\documentclass[11pt]{article}

\usepackage[letterpaper, text={6.5in, 9.0in}]{geometry}
\usepackage{authblk}

\RequirePackage{amsthm,amsmath,amsfonts,amssymb,mathrsfs}
\RequirePackage[authoryear]{natbib}
\RequirePackage[colorlinks,citecolor=blue,urlcolor=blue]{hyperref}
\RequirePackage{graphicx}
\usepackage{multirow}
\usepackage{booktabs}
\usepackage{bbm}
\RequirePackage{tikz}
\usetikzlibrary{decorations.pathreplacing}

\usepackage{booktabs}

\theoremstyle{plain}

\newtheorem{theorem}{Theorem}

\newtheorem{proposition}{Proposition}
\theoremstyle{definition}
\newtheorem{definition}{Definition}
\newtheorem{assumption}{Assumption}

\def\real{\mathbb{R}}
\def\indpt{\perp\!\!\!\!\perp}
\def\t{\top}
\def\H{\mathcal{H}}
\def\F{\mathcal{F}}
\def\f{f}
\def\G{\mathcal{G}}
\def\ee{E}
\def\vv{\mathrm{var}}
\def\enc{Q}
\def\net{h}
\def\mdd{r} 
\def\L{\mathcal{L}}
\def\fpca{\varPhi }
\def\kk{\mathrm{K}}
\def\T{\varPi}
\def\netclass{H_{\mathrm{NN}}}
\def\loss{\mathcal{E}}

\begin{document}

\title{Deep Nonlinear Sufficient Dimension Reduction for Discretely Observed Functional Data}

\author[1]{Baiyu Chen}
\author[2]{Jun Fan}
\author[3]{Zhou Yu}

\affil[1]{Faculty of Business for Science and Technology, School of Management, University of Science and Technology of China}

\affil[2]{Department of Mathematics, Hong Kong Baptist University}

\affil[1]{School of Statistics, East China Normal University}

\date{}
\maketitle

\begin{abstract}
We develop a modular framework for deep nonlinear Sufficient Dimension Reduction (SDR) of discretely observed functional predictors with Euclidean or metric-space-valued responses. A presmoothing encoder converts the functional input into finite-dimensional coordinates, and a neural network learns a sufficient representation from them. The framework separates the choice of encoder from that of the SDR criterion. We study kernel embedding and discrete functional principal component analysis, each combined with either a generalized martingale difference divergence network or a Fr\'echet cumulative covariance network. These combinations share objectives of the martingale difference divergence form. {Under input smoothness and general H\"older regularity of the target operator, we establish approximation and generalization bounds that jointly account for functional truncation, network approximation, statistical estimation, numerical integration, measurement noise, and encoder estimation. The complete theoretical analysis rigorously demonstrates how errors in the encoded coordinates propagate through the network to the deep nonlinear SDR estimator, and quantifies the final estimation error in relation to observation frequency and model complexity.} Simulations and an application to satellite-derived vegetation trajectories and crop-composition responses illustrate the statistical and computational advantages over RKHS-based nonlinear functional SDR methods.  

\bigskip 
\noindent
{\bf Keywords}: 
Sufficient dimension reduction,
sliced inverse regression,
sliced average variance estimation,
functional data,
metric space,
neural network.

\end{abstract}

\section{Introduction}

A central challenge in analyzing functional data is to reduce an entire trajectory to a few features that retain its information about a response. Nonlinear functional sufficient dimension reduction (SDR) formalizes this goal through conditional independence. For a functional predictor $X$ taking values in a function space $\H$ and a response $Y$, we seek a nonlinear map $f:\H\to\real^d$, with a small integer $d$, such that
\begin{equation}\label{2}
Y\indpt X\mid f(X).
\end{equation}
The representation $f(X)$ then contains all the information in $X$ relevant to $Y$, while allowing subsequent analysis to proceed in a low-dimensional space. In practice, however, the trajectory $X$ is usually available only through noisy measurements at finitely many locations, which may differ across subjects. Learning $f$ therefore requires connecting a target defined on whole functions with the discrete observations actually available.

For an $\real^p$-valued predictor $X$, classical linear SDR restricts the representation to $f(X)=\beta^\t X$ with $\beta\in\real^{p\times d}$ and seeks the associated central subspace. Representative methods include sliced inverse regression \cite{Li1991}, sliced average variance estimation \cite{CookWeisberg1991}, contour regression \cite{LiZhaChia2005}, directional regression \cite{LiWang2007}, minimum average variance estimation \cite{Xia2002} and semiparametric methods \cite{MaZhu2013,MaZhu2014}. Two lines of work extend this formulation by relaxing different restrictions. Functional linear SDR accommodates curves by replacing vector projections with inner products between functions \cite{FerreYao2003,FerreYao2005,HsingRen2009}, including settings with sparse observations \cite{Yao2015}. Nonlinear SDR for vector-valued predictors instead relaxes the linear form of the reduction and captures more general predictor--response relationships \cite{LiArteLi2011,LeeLiChia2013,TangLi2026}. It is then obvious that the problem (\ref{2}) brings together above two major extensions of classical SDR. 

{Li and Song \cite{LiSong2017} first developed a general theory for nonlinear functional SDR (\ref{2}) through nested reproducing kernel Hilbert spaces (RKHSs). Their construction provides a foundation for the problem, but several limits remain. The second-level RKHS is determined by the first-level one, so the modeling of sufficient predictors is to some extent, constrained by curve modeling, compromising its flexibility. The convergence theory for partially observed curves only assumes an estimation rate $\delta_n$ for the RKHS covariance operators, without deriving this rate for specific observation schemes and smoothing procedures. The dependence of the final estimation error on the observation process is therefore left implicit in $\delta_n$. Finally, the implementation of kernel methods involves sample-sized Gram matrices, making computation demanding as the sample grows.}

Recent neural approaches offer another way to represent nonlinear sufficient predictors. For vector-valued inputs, dependence-based objectives have been used to train neural networks for SDR \cite{Huang2024,Chen2024,Yuan2026}, yielding representations that can be evaluated through the fitted network without referring to the training observations. Extending this idea to functional predictors requires an additional step: a standard feedforward network takes a finite-dimensional vector as input, whereas the domain of the desired operator is a function space. {Drawing on operator learning \cite{ChenChen1995,Ana2020,Lanthaler2023,Lu2021,FanZhou2026}, we therefore develop a general modular encoder--network architecture, $f=h\circ Q$, where an encoder $Q:\H\to\real^r$ supplies finite-dimensional coordinates of the functional predictor and a neural network $h:\real^r\to\real^d$ learns a sufficient representation from them. The encoder is designed to process the functional structure of $X$ from its discrete observations, while the network is trained with a response-aware SDR objective to enforce the desired conditional independence. Neither choice determines the other.}

In other words, the newly proposed framework allows for separate choices of function representation and low-dimensional feature extraction criteria for SDR, and accommodates various encoders and dimension reduction methods. In this paper, we study kernel embedding and discrete functional principal component analysis (discrete FPCA) as two presmoothing encoders within this architecture. Kernel embedding applies a prescribed integral smoothing operator to each input function and then projects the result onto a finite-dimensional basis \cite{Shi2025}. Discrete FPCA instead pools observations across subjects to smooth the covariance function and constructs coordinates from its estimated eigenfunctions, following the covariance-based approach to sparsely observed functional data \cite{Yao2005}. These procedures allow irregular measurements by smoothing either individual inputs or the shared covariance structure. Either encoder is then paired with a generalized martingale difference divergence network (GMDD-Net; \cite{Chen2024}) for Euclidean responses, or a Fr\'echet cumulative covariance network (FCCov-Net; \cite{Yuan2026}) for metric-space-valued responses such as distributions, positive definite matrices, and manifold-valued observations. Both objectives have the martingale difference divergence (MDD; \cite{ShaoZhang2014}) form, making it possible to analyze the resulting estimators together.

{The main theoretical contribution is the establishment of a comprehensive, end-to-end analytical framework. Our theory rigorously links the discrete nature of observations to the final nonlinear functional SDR estimation risk by providing explicit error bounds. This is achieved by systematically deriving and integrating the observation, encoder, approximation, and statistical errors, offering a holistic view of the estimator's performance.} Our theory first identifies the approximation error that remains even with complete observations and a known encoder. A finite-dimensional encoder discards part of the functional input, and the network introduces a further approximation error on the retained coordinates. We study these two effects jointly under smoothness assumptions on the input and general H\"older regularity of the target operator. For kernel embedding, input regularity is expressed through deterministic Besov or Gaussian RKHS conditions, leading to uniform approximation bounds. For FPCA, it is expressed through covariance eigenvalue decay, leading to bounds in expectation. Increasing the encoder dimension reduces the truncation error but raises the dimension of the function that the network must approximate. The approximation bounds therefore identify how encoder dimension and network size should be balanced. They range from logarithmic to faster subpolynomial rates as input regularity strengthens. Throughout, we retain a general H\"older class of target operators, rather than imposing additional analytic or problem-specific structure solely to obtain polynomial rates. This scope is appropriate for SDR, where the dimension reduction map is determined by a model--free predictor--response relationship.

To pass from this ideal approximation problem to the observed data, we bound the discrepancies between population coordinates and those obtained from the discrete noisy measurements. Both encoders incur numerical integration error and measurement noise; discrete FPCA introduces an additional encoder estimation error and its control builds on the functional data analysis literature. Covariance smoothing theory describes the effects of observation frequency and transitions between sparse and dense regimes \cite{Yao2005,ZhangWang2016,ZhouYaoZhang2023,JLQi2025,ShaoLinYao2022}, while Zhou, Wei and Yao \cite{ZhouWeiYao2025} treated diverging eigenfunction indices. {After obtaining nonasymptotic expected bounds for estimated eigenfunctions in discrete FPCA that are not available from existing PCA-based operator learning literature \cite{Lanthaler2023,Liu2023}, we then control how coordinate errors propagate through the network and into the MDD objectives, accounting for their effects on both training and population risk. Combing these bounds with the approximation and empirical-process terms gives explicit learning rates in terms of the number of curves, measurements per curve, and model complexity. The resulting parameter choices balances the benefit of retaining more coordinates ans using a larger network against the accumulated observation and encoder estimation errors. }

{Lastly, another obvious advantage of our proposed framework is computation efficiency, compared with kernel methods in Li and Song \cite{LiSong2017}. The training and inference time scales well with the sample size and the training data need not be referenced when computing the representations of unseen instances.} Our simulations show that when the sample size is 3000, the average runtimes of f-GSIR and f-GSAVE in \cite{LiSong2017} are approximately 2.48 and 4.31 times that of kernel embedding with GMDD-Net, respectively.

The rest of the paper is organized as follows. Section 2 defines the central class of functional nonlinear SDR, and Section 3 presents the encoder--network architecture and its MDD-form objectives. Section 4 introduces two presmoothing encoders and their approximation results. Section 5 establishes generalization bounds of the SDR operators, including the nonasymptotic analysis of discrete FPCA. Sections 6 and 7 present simulation studies and a real-world data application, respectively. Section 8 concludes this paper. Proofs and additional numerical results are provided in the Supplement.

Throughout the paper, for a Borel measure $\mu$ on $[0,1]$, let $L_2(\mu)$ denote the space of square-integrable functions with respect to $\mu$, equipped with inner product $\langle\cdot,\cdot\rangle_{L_2(\mu)}$ and norm $\|\cdot\|_{L_2(\mu)}$. For Lebesgue measure, we simply write $L_2([0,1])$, with inner product $\langle\cdot,\cdot\rangle_{L_2}$ and norm $\|\cdot\|_{L_2}$. We use $X$ for a random function and $x$ for a deterministic element of this space. We write $A\lesssim B$ if $A\leq CB$ for some positive constant $C$, and $A\asymp B$ if both $A\lesssim B$ and $B\lesssim A$ hold. Unless stated otherwise, implicit constants may depend on fixed model parameters but are independent of sample sizes and other varying quantities. Positive constants $c$ and $C$ may change from one occurrence to another.

\section{Functional nonlinear SDR}

For completeness, we recall the functional nonlinear SDR $\sigma$-fields and classes introduced in \cite{LeeLiChia2013,LiSong2017}. Let $(\Omega, F, P)$ be a probability space, $H_X$ a Hilbert space of functions from $[0,1]$ to $\real$, and $H_Y$ a metric space with metric $d_Y$. In this paper, we focus on $H_X\subset L_2([0,1])$ and $H_Y=\real^q$. It will be shown later that our development also accommodates $H_Y$ that is non-Euclidean.

Let $F_X$ be the Borel $\sigma$-field induced by the inner product on $H_X$, and let $F_Y$ be the Borel $\sigma$-field on $H_Y$. Define $X:\Omega\rightarrow H_X$ and $Y:\Omega\rightarrow H_Y$ as random elements measurable with respect to $F/F_X$ and $F/F_Y$, respectively.

Let $P_X$ and $P_Y$ be the distributions of $X$ and $Y$, respectively; that is, the induced measures $P\circ X^{-1}$ and $P\circ Y^{-1}$ on $(H_X,F_X)$ and $(H_Y,F_Y)$. We take the pair of random elements to have values in $(H_X\times H_Y,F_X\times F_Y)$ and write $P_{XY}=P\circ(X,Y)^{-1}$ for the distribution of $(X,Y)$. Let $\sigma(X)$ be the $\sigma$-field in $F$ generated by $X$, i.e., $\sigma(X)=X^{-1}(F_X)$. Denote the conditional distribution of $X$ given $Y$ by $P_{X\mid Y}:H_Y\times F_X\rightarrow\real$.

\begin{definition}
    Assume that the family of probability measures $\{P_{X\mid Y}(\cdot\mid y):y\in H_Y\}$ is dominated by a $\sigma$-finite measure. By Theorem 1 of \cite{LeeLiChia2013}, there exists a unique sub-$\sigma$-field $\G_{Y\mid X}$ of $\sigma(X)$ such that
\begin{equation}\label{3}
Y\indpt X\mid \G_{Y\mid X}.
\end{equation}
The field $\G_{Y\mid X}$ is called the central $\sigma$-field and is the target of functional nonlinear SDR. Every sub-$\sigma$-field satisfying (\ref{3}) contains $\G_{Y\mid X}$, i.e., $\G_{Y\mid X}$ is minimal.
\end{definition}

The central $\sigma$-field describes functional dimension reduction solely in terms of conditional independence and rather abstract to handle. To work with explicit functions, we introduce square integrability, which also allows us to use orthogonality and projection. Let $\L_2(P_X)$ denote the class of functions $\f$ satisfying $\ee \f^2(X)<\infty$ under $P_X$. Elements of $\L_2(P_X)$ are defined on the function space $H_X$, rather than on [0,1], which motivates the distinct notation for this $L_2$ space. Define $\L_2(P_Y)$ and $\L_2(P_{XY})$ analogously. For a sub-$\sigma$-field $\G$ of $\sigma(X)$, let $\mathfrak{M}_{\G}$ be the class of functions $\f$ in $\L_2(P_{X})$ for which $\f(X)$ is $\G$-measurable.

\begin{definition}
    Let $\G$ be an SDR $\sigma$-field and $\G_{Y\mid X}$ the central $\sigma$-field. Then $\mathfrak{M}_{\G}$ is called an SDR class, and $\mathfrak{M}_{\G_{Y\mid X}}$ is called the central class that is denoted by $\mathfrak{S}_{Y\mid X}$.
\end{definition}

The central class consists of square-integrable functions measurable with respect to the central $\sigma$-field $\G_{Y\mid X}$ and generalizes the central subspace of linear SDR.

\section{Deep functional nonlinear SDR architecture}
The central class $\mathfrak{S}_{Y\mid X}$ is still abstract and now we turn from the population target to a working class of operators that can be fitted from data. To represent the central class $\mathfrak{S}_{Y\mid X}$, \cite{LiSong2017} used a nested RKHS. We adopt an encoder--network architecture from operator learning, separating the representation of the functional input from the supervised learning of its nonlinear reduction.

To represent the nonlinear SDR operator $\f$ in (\ref{2}), we first connect the functional input to a finite-dimensional network. An encoder $\enc$ generates the vector of coordinates passed to the network $\net$, giving the representation
\begin{equation*}
f=\net\circ \enc
\end{equation*}
where $\circ$ denotes function composition. The encoder $\enc$ summarizes the functional input $X$ as an $\mdd$-dimensional vector, where $\mdd>d$ is an integer. The map $\net :\real^\mdd\rightarrow \real^d$ is a feedforward network with the rectified linear unit (ReLU) activation function.

We consider two choices for the encoder $\enc$: \emph{kernel embedding}, which is specified independently of the sample, and \emph{functional principal component analysis (FPCA)}, whose basis needs to be estimated from the data. Both use presmoothing to construct a common representation from discrete observations. Kernel embedding smooths each input through an integral operator, whereas discrete FPCA estimates a smooth covariance surface and represents trajectories in its leading eigendirections. They will be detailed in subsequent sections.

Write $\F=\{f: f=\net\circ\enc\}$ for this operator class. To relate this working representation to the central class, we require $\F$ to be sufficiently rich. Following \cite{LiSong2017}, we impose the following condition.

\begin{assumption}\label{ass_class}
The operator class $\F$ is a dense subset of $\L_2(P_X)$ modulo constants, i.e., for any $f\in \L_2(P_X)$, there is a sequence $\{f_n\}\subseteq \F$ such that $\vv\{f_n(X)-f(X)\}\rightarrow 0$.
\end{assumption}

Under Assumption \ref{ass_class}, the operators in $\F$ that are $\G_{Y\mid X}$-measurable constitute the central class $\mathfrak{S}_{Y\mid X}$, so we restrict attention to operators of the form $\net\circ \enc$. A subspace $\mathfrak{S}$ of $\F$ is called unbiased if it is contained in $\mathfrak{S}_{Y\mid X}$, and exhaustive if it equals $\mathfrak{S}_{Y\mid X}$ \cite{LiSong2017}.

\subsection{GMDD-Net}
The encoder $\enc$ is constructed from the predictor alone. A response-based objective is therefore needed to train the network $\net$ toward the SDR target $Y\indpt X\mid \net\circ\enc(X)$. GMDD-Net \cite{Chen2024} gives such an objective for $X\in\real^{p}$. We extend this construction to function $X$ by applying the same dependence criterion to the output of the encoder--network composition. Generalized martingale difference divergence (GMDD; \cite{Lilu2023}) measures conditional mean dependence of a random vector $U\in\real^q$ on a random vector $V\in\real^p$. It generalizes MDD and is defined by
\begin{equation*}
\mathrm{GMDD}(U\mid V)=-\ee\{(U-\ee U)^\t(\widetilde{U}-\ee\widetilde{U})\kappa(V,\widetilde{V})\},
\end{equation*}
where $(\widetilde{U},\widetilde{V})$ is an independent copy of $(U,V)$ and $\kappa(v_1,v_2)=\int_{\real^p\setminus \{0\}}[1-\cos\{s^\t(v_1-v_2)\}]d\mu(s)$, with $\mu$ a L\'evy measure; see \cite{Chen2024} for examples. Moreover, $\mathrm{GMDD}(U\mid V)\geq 0$, with $\mathrm{GMDD}(U\mid V)=0$ if and only if $\ee(U\mid V)=\ee(U)$ almost surely.

Since $\mathfrak{S}_{Y\mid X}$ is a closed subspace of $\L_2(P_X)$, the space $\L_2(P_X)$ admits the orthogonal decomposition $\L_2(P_X)=\mathfrak{S}_{Y\mid X}\oplus \mathfrak{S}_{Y\mid X}^\perp$, where $\mathfrak{S}_{Y\mid X}^\perp$ is the orthogonal complement of $\mathfrak{S}_{Y\mid X}$. Let $f_0\in\mathfrak{S}_{Y\mid X}$. For any function $g\in\mathfrak{S}_{Y\mid X}^\perp$, we have
\begin{equation*}
\ee\{g(X)\mid Y\}=\ee[\ee\{g(X)\mid f_0(X),Y\}\mid Y]=\ee[\ee\{g(X)\mid f_0(X)\}\mid Y]\overset{a.s.}{=}\ee\{g(X)\},
\end{equation*}
where the second equality follows from $Y\indpt X\mid \mathfrak{S}_{Y\mid X}$ and the third from $g\in\mathfrak{S}_{Y\mid X}^\perp$. Thus, for any $g$ in the orthogonal complement of $\mathfrak{S}_{Y\mid X}$, $\mathrm{GMDD}(g(X)\mid Y)=0$. Since GMDD is nonnegative, maximizing it over $f\in \L_2(P_X)$ drives $f$ away from the orthogonal complement space $\mathfrak{S}_{Y\mid X}^\perp$, thereby pulling it toward $\mathfrak{S}_{Y\mid X}$.

With $f$ represented by $h\circ Q$, we use GMDD to train $h$. In practice, the target dimension $d$ may be unknown, so we initially use a working dimension $m$; dimension selection is described in Section~\ref{determine}. Given i.i.d. observations $(X_i,Y_i),i=1,...,n$, we first construct the empirical encoder $\widehat{\enc}$ and then train a GMDD-Net $\widehat{\net}$ by
\begin{equation}\label{gmdd}
    \widehat{\net}\in\mathop{\arg\min}_{\net\in \netclass} -\mathrm{GMDD}_n(\net\circ\widehat{\enc}(X)\mid Y)+\eta\|\vv_n\{\net\circ\widehat{\enc}(X)\}-I_m\|_F^2,
\end{equation}
where $\netclass$ is a ReLU network class and $\mathrm{GMDD}_n(\cdot\mid \cdot)$ is an unbiased fourth-order U-statistic estimator of GMDD, detailed as (34) in the supplement. The regularization term forces the variance of $h\circ \widehat{Q}(X)$ to be identity, where $\vv_n(\cdot)$ denotes the empirical covariance matrix, $\|\cdot\|_F$ the Frobenius norm, and $\eta$ the regularization parameter. When the encoder must be estimated, its estimation and the subsequent network training both use the observed sample. Some approaches, such as \cite{Liu2023}, split the sample into disjoint subsets for these two stages. Although this simplifies the analysis, it reduces the sample size available to each stage. We use the full sample for both stages, a choice also retained in the generalization analysis. When the encoder $\enc$ is independent of the sample, replace $\widehat{\enc}$ by the original $\enc$.

\subsection{Accommodating metric-space-valued \texorpdfstring{$Y$}{Y}}
To accommodate responses in a metric space $H_Y$ other than $\real^q$, we change the dependence criterion used to train the network. \cite{Yuan2026} introduced FCCov-Net for nonlinear SDR with $X\in\real^p$ and $Y\in H_Y$, where $d_Y$ is the response metric. Its formulation makes this extension possible within the same architecture. For a random element $V\in H_Y$ and a random variable $U\in\real$ satisfying $\vv(U)<\infty$ and $\ee d_Y^2(V,\widetilde{V})<\infty$, the Fr\'echet cumulative covariance is defined by
\begin{equation*}
    \mathrm{FCCov}(U\mid V)=\ee\big(\mathrm{cov}^2[U,\mathbbm{1}\{d_Y(V,\bar{V})<d_Y(V,\widetilde{V})\}\mid \bar{V},\widetilde{V}] \big),
\end{equation*}
where $\mathbbm{1}(\cdot)$ denotes the indicator function. As with GMDD, $\mathrm{FCCov}(U\mid V)\geq0$, and $\mathrm{FCCov}(U\mid V)=0$ if and only if $\ee(U\mid V)=\ee(U)$ almost surely. The same argument shows that any $g$ in $\mathfrak{S}_{Y\mid X}^\perp$ satisfies $\mathrm{FCCov}(g(X)\mid Y)=0$. This motivates estimating $f$ in $\mathfrak{S}_{Y\mid X}$ by training an FCCov-Net $\widehat{h}$ according to
\begin{equation}\label{fccov}
   \widehat{\net}\in\mathop{\arg\min}_{\net\in \netclass}-\mathrm{FCCov}_n(\net\circ\widehat{\enc}(X)\mid Y)+\eta\|\vv_n\{\net\circ\widehat{\enc}(X)\}-I_m\|_F^2,
\end{equation}
where $\mathrm{FCCov}_n(\cdot\mid\cdot)$ is also a fourth-order U-statistic (see \cite{Yuan2026}). The conditional-mean-dependence characterization of FCCov gives the same population motivation as GMDD while allowing non-Euclidean $Y$. We therefore state the subsequent theoretical results for GMDD-Net; the same conclusions apply to FCCov-Net.

Both GMDD and FCCov are induced by MDD, providing two related criteria for the supervised stage. The two stages play distinct roles: kernel embedding and discrete FPCA handle the functional input, while GMDD-Net and FCCov-Net specify how the response enters the reduction. Either presmoothing encoder can be paired with either network objective, with the response space guiding the choice.

The following theorem confirms the validity of (\ref{gmdd}) and (\ref{fccov}) by a population-level statement.

\begin{theorem}\label{unbiased}
Let $m$ be a positive integer no greater than the dimension of $\mathfrak{S}_{Y\mid X}$, which may be infinite. If $f_0=\big(f_0^{[1]},...,f_0^{[m]}\big)$ solves the optimization problem
\begin{equation*}
    \max_{f\in\{\L_2(P_X)\}^m} \left\{
        \begin{aligned}
        \mathrm{GMDD}[f(X)\mid Y], & \quad H_Y \text{ Euclidean}\\
        \mathrm{FCCov}[f(X)\mid Y], &\quad H_Y \text{ metric space}
        \end{aligned}
    \right.,\text{ s.t. } \mathrm{var}[f(X)]=I_m.
\end{equation*}
then $f_0^{[i]}\in\mathfrak{S}_{Y\mid X}$ for all $1\leq i\leq m$.
\end{theorem}

Theorem~\ref{unbiased} establishes that $\mathrm{span}(f_0^{[1]},...,f_0^{[m]})$, obtained by optimizing GMDD or FCCov, is unbiased for the central class $\mathfrak{S}_{Y\mid X}$. The proof follows from Theorem 3.2 of \cite{Chen2024}, whose argument applies both to $f$ defined on $\real^p$ and to $f$ defined on a function space.

\section{Two encoders and their approximation results}
The preceding section specifies how the response guides the learned reduction. We now examine how the choice of encoder affects the approximation of the target operator. 

\subsection{Kernel embedding}\label{kernel_app}
\subsubsection{Method}
Kernel embedding \cite{Smola2007,Shi2025} first smooths an input function through an integral operator and then projects the result onto a finite basis. This presmoothing step accommodates discrete observations without requiring a common observation grid. We describe this encoder for a deterministic input $x\in H_X$ and subsequently establish approximation bounds uniformly over $H_X$. The kernel embedding of $x$ is
\begin{equation*}
    L_Kx=\int_{0}^1K(\cdot,t)x(t)dt,
\end{equation*}
where $L_K$ is the integral operator induced by a continuous, symmetric, positive-definite Mercer kernel $K:[0,1]\times [0,1]\rightarrow\real$. The embedded function $L_Kx$ is then mapped by the projection operator $\T$ to its coefficients in the basis $\{\psi_i\}_{i=1}^\mdd$:
\begin{equation*}
\T(L_Kx)=\left(\int_0^1(L_Kx)(s)\psi_1(s)d\mu(s),...,\int_0^1(L_Kx)(s)\psi_\mdd(s)d\mu(s)\right)^\t,
\end{equation*}
where $\{(\nu_i,\psi_i)\}$ is the eigensystem of a second integral operator $L_{K_0}^\mu$, induced by a Mercer kernel $K_0$ and a positive Borel measure $\mu$ on $[0,1]$: $L_{K_0}^\mu x=\int_{0}^1K_0(\cdot,t)x(t)d\mu(t)$. The compactness, self-adjointness and positivity of $L_{K_0}^\mu$ ensure that $\{\psi_i\}$ is well defined by the spectral theorem \cite{HsingEubank2015}.

The smoothing and projection kernels determine the representation supplied to the network. For the approximation results below, we use Gaussian kernels and their linear combinations. Define the Gaussian kernel
\begin{equation*}
k_\gamma(s,t)=\exp\{-(s-t)^2/\gamma^2\},\quad s,t\in\real,
\end{equation*}
and a linear combination of Gaussian kernels:
\begin{equation*}
K_{\varrho,\gamma}(s,t)=\sum_{j=1}^\varrho\binom{\varrho}{j}(-1)^{1-j}\frac{1}{j}\left(\frac{1}{\gamma^2\pi}\right)^{\frac{1}{2}}k_{j\gamma}(s,t),\quad s,t\in\real,
\end{equation*}
where $\varrho=\lfloor\alpha\rfloor+1$ depends on the smoothness parameter $\alpha$ of the input function space $H_X$ (see Assumption~\ref{F_modulus}). Choosing the embedding kernel $K$ as $K(s,t)=k_\gamma(s,t)u(t)$, with $u$ the density of a positive Borel measure $\mu$, gives $L_Kx\in H_\gamma([0,1])$ for any $x\in L_2([0,1])$, where $H_\gamma([0,1])$ is the RKHS induced by $k_\gamma$. Choosing $K$ as $K(s,t)=k_{\varrho,\gamma}(s,t)$ also gives $L_Kx\in H_\gamma([0,1])$ for any $x\in L_2([0,1])$. The embedding and projection kernels $K$ and $K_0$ may differ, and kernel parameters such as $\gamma$ can be tuned. This flexibility allows the encoder to exploit input regularity and achieve improved approximation rates.

\subsubsection{Approximation rates}
Now the nonlinear functional SDR operator takes the form $f=\net\circ (\T\circ L_K)$, where $\net$ is a GMDD-Net. We next study the approximation properties of this architecture.

We impose regularity conditions on the input function space $H_X$ and the target operator $f: H_X\rightarrow\real^d$. These conditions serve distinct roles: input smoothness controls the error introduced by encoding, while continuity of the target controls how errors in the input affect its output. Since $L_2([0,1])$ is too general to establish approximation theories, we assume that $H_X$ lies in the Besov space $B_{2,\infty}^\alpha([0,1])\subset L_2([0,1])$, whose parameter $\alpha$ measures smoothness. See Appendix A.1 in the Supplement for definitions of Besov and Sobolev spaces.

Having specified input smoothness, we describe the regularity of the target operator. Write its $i$th component as $f_i:H_X\rightarrow\real,i=1,...,d$. The modulus of continuity of $f_i$ is
\begin{equation*}
\omega_{f_i}(t)=\sup\{|f_i(x_1)-f_i(x_2)|: x_1,x_2\in H_X,\|x_1-x_2\|_{L_2(\mu)}\leq t|\},
\end{equation*}
where $\mu$ is a positive Borel measure, such as Lebesgue measure or a Gaussian measure restricted to $[0,1]$.

\begin{assumption}\label{F_modulus}
(i) $H_X\subset  B_{2,\infty}^\alpha([0,1])$. (ii) Each component of the target operator $f$, denoted by $f_i$, is continuous, and its modulus of continuity $\omega_{f_i}(t)$ satisfies $\omega_{f_i}(t)\leq C_f \cdot t^{\lambda}$ for some $\lambda\in (0,1]$ and positive constant $C_f$.
\end{assumption}

An approximation rate for the kernel embedding--network architecture when the input space is a subset of $B_{2,\infty}^\alpha(\real)$ is established in \cite{Shi2025}. As noted there, this result does not directly apply to $B_{2,\infty}^\alpha(\mathcal{S})$ for any subset $\mathcal{S}$ of $\real$. Our predictors are defined on [0,1], so we first construct a bounded extension from $B_{2,\infty}^\alpha([0,1])$ to $B_{2,\infty}^\alpha(\real)$. This allows the whole-line approximation framework to be used for inputs on the observed interval. The following theorem therefore uses the embedding kernel $K_{\mathfrak{E},\varrho,\gamma}$ obtained by combining this extension with $K_{\varrho,\gamma}$. Appendix C in the Supplement gives the construction and an explicit formula for $K_{\mathfrak{E},\varrho,\gamma}$ (see (12) there). For the projection operator $L_{K_0}^\mu$, we take $K_0=k_\gamma$ and choose $\mu$ as the Gaussian measure with density $u(t)=\frac{\mathrm{b}}{\sqrt{\pi}}\exp(-\mathrm{b}^2 t^2)$, where $\mathrm{b}$ is a scaling parameter.

\begin{theorem}\label{kernel_appro_1}
Suppose that Assumptions~\ref{ass_class}, \ref{F_modulus} hold and that $\|x\|_{B_{2,\infty}^\alpha}\leq 1$ for every $x\in H_X$. Let $K=K_{\mathfrak{E},\varrho,\gamma}$ be the embedding kernel and $K_0=k_\gamma$ the projection kernel associated with a Gaussian measure $\mu$. For suitable choices $\gamma\asymp\frac{(\log\log M)^2}{\log M}$ and $\mdd\asymp\frac{\log M}{\log\log M}$, there exists a neural operator $f_{\mathrm{NN}}=h_{\mathrm{NN}}\circ\T\circ L_{K_{\mathfrak{E},\varrho,\gamma}}$, where $\net_{\mathrm{NN}}$ is a finite-dimensional deep ReLU network with $M$ nonzero parameters and depth $L\lesssim\big(\frac{\log M}{\log\log M}\big)^2$, such that
\begin{equation*}
\sup_{x\in H_X}\|f_{\mathrm{NN}}(x)-f(x)\|_2\lesssim (\log M)^{-\alpha\lambda}(\log\log M)^{2\alpha\lambda}.
\end{equation*}
\end{theorem}

The embedded functions $L_Kx\in H_\gamma([0,1])$ satisfy $\sup_{x\in H_X}\|L_Kx\|_{H_\gamma}\leq C\gamma^{-1/2}$ for some constant $C>0$. This bound has the spectral form
\begin{equation}\label{source1}
\sup_{x\in H_X}\sum_{j\geq 1}\frac{|\langle x,\psi_j\rangle_{L_2(\mu)}|^2}{\nu_j}\leq C\gamma^{-1},
\end{equation}
where $\{(\nu_j,\psi_j)\}$ are the eigenpairs of $L_{K_0}^\mu$. This is an ellipsoid condition: the RKHS norm is a weighted $\ell^2$ norm of the coefficients in the Mercer eigenbasis; \cite{Bach2017} gives the same spectral interpretation of RKHS norms via inverse eigenvalue weights. 

To see how the two encoder steps interact with network approximation, decompose the error in Theorem~\ref{kernel_appro_1} into three terms:
\begin{equation}\label{decom1}
    \begin{split}
    \|f_{\mathrm{NN}}(x)-f(x)\|_2 \leq &\|f(x)-f(L_Kx)\|_2+\|f(L_Kx)-f\{\varPi^{-1}\circ\varPi(L_K x)\}\|_2\\
    &+\|f\circ\varPi^{-1}\{\varPi(L_Kx)\}-f_{\mathrm{NN}}(x)\|_2.
    \end{split}
\end{equation}
The first term is the embedding error from approximating $x$ by $L_Kx$. The second is the projection error for $L_Kx$, where $\varPi^{-1}$ is the inverse map of $\varPi$. The third is the error of approximating a $\lambda$-H\"older continuous function by a deep ReLU network. The bounds for all three terms depend on $\gamma$. Tuning $\gamma$ therefore balances the quality of the encoded representation against the subsequent network approximation, and this balance determines the resulting rate. This is the role of the flexibility in the smoothing kernel noted above.

The rate in Theorem~\ref{kernel_appro_1} is polynomial in $\log M$, rather than in $M$, reflecting the difficulty of approximation on an infinite-dimensional input space. For vector inputs $x\in[0,1]^p$ with fixed $p$, a ReLU network can approximate a smooth function $f:[0,1]^p\rightarrow\real$ at a polynomial rate in $M$, roughly $M^{-s/p}$ \cite{Yaro2018,QiWang2022,QiWang2024}. For functional inputs $x\in H_X\subset L_2([0,1])$, the encoder first represents $x$ by $\mdd$ coordinates. To improve accuracy over a Sobolev or Besov class, the retained dimension $\mdd$ must increase. Retaining more coordinates reduces the information loss by the encoder, but also increases the dimension of the function that the network must approximate. The bound reflects this balance. We next examine how it changes when stronger input smoothness allows a more accurate representation with fewer coordinates. Specifically, $H_\gamma([0,1])$ is the Gaussian RKHS induced by $k_\gamma$ and is contained in $B_{2,\infty}^\alpha([0,1])$.

\begin{theorem}\label{kernel_appro_2}
Suppose that Assumptions~\ref{ass_class}, \ref{F_modulus} hold. Further assume that $H_X$ is a subset of the unit ball in $H_\gamma([0,1])$. Let $K_0=k_\gamma$ be the projection kernel associated with a Gaussian measure $\mu$, and let $K(s,t)=k_\gamma(s,t)u(t)$ be the embedding kernel, where $u(t)$ is the density of $\mu$. For a suitable choice $\mdd\asymp(\log M)^{1/2}$, there exists a neural operator $f_{\mathrm{NN}}=h_{\mathrm{NN}}\circ\T\circ L_{K}$, where $\net_{\mathrm{NN}}$ is a finite-dimensional deep ReLU network with $M$ nonzero parameters and depth $L\lesssim\log M$, such that
\begin{equation*}
\sup_{x\in H_X}\|f_{\mathrm{NN}}(x)-f(x)\|_2\lesssim \exp\{-\sqrt{1/3}\lambda(\log M)^{1/2}\}(\log M)^{1/2}.
\end{equation*}
\end{theorem}

The error decomposition for Theorem~\ref{kernel_appro_2} follows (\ref{decom1}), with the embedding-error term absent because $x$ already belongs to an RKHS. The assumption $H_X\subset H_\gamma([0,1])$ means
\begin{equation}\label{source2}
\sum_{j\geq1}\frac{|\langle x,\psi_j\rangle_{L_2(\mu)}|^2}{\nu_j}\leq 1,
\end{equation}
where $(\nu_j,\psi_j)$ are again the eigenpairs of $L_{K_0}^\mu$. This is also an ellipsoid condition. Unlike the $\gamma$-dependent condition (\ref{source1}), (\ref{source2}) is fixed and more restrictive. This also explains why we need not tune $\gamma$ in Theorem~\ref{kernel_appro_2}. 

The dominant term in Theorem~\ref{kernel_appro_2} is $\exp\{-\sqrt{1/3}\lambda(\log M)^{1/2}\}$, equivalently $M^{-c/(\log M)^{1/2}}$ for some constant $c>0$. This decays faster than $(\log M)^{-l}$ for every $l>0$, but more slowly than $M^{-l}$, a polynomial rate in $M$. Thus, even with stronger input smoothness, the bound obtained here remains subpolynomial in $M$. Polynomial rates may nevertheless be achievable under additional structural assumptions on the target operator, such as  analytical/holomorphic dependence \cite{Schwab2019,Herrmann2024}, low intrinsic dimension \cite{Kutyniok2022,Liu2023} and Barron-type structure \cite{YangXiang2022}. Such rates have also been established for certain PDE solution operators by exploiting their specific structures \cite{Lanthaler2023}. However, to maintain methodological generality, we do not impose these additional assumptions in this paper.

\subsection{FPCA}
\subsubsection{Method}
We next consider an encoder that adapts its coordinates to the variation in the functional predictor. The FPCA basis is determined by the data distribution and must be estimated in practice. Its dependence on the input distribution also changes the natural formulation of the approximation results. Kernel embedding uses deterministic smoothness assumptions and uniform bounds over an input function space. FPCA depends on the distribution of $X$ through its covariance operator, so we explicitly treat $X$ as random. Its regularity assumptions concern distributional quantities, particularly covariance eigenvalue decay, and its approximation bounds hold in expectation with respect to $X$.

Assume without loss of generality that $\ee X=0$, and suppose $\ee \|X\|_{L_2}^2<\infty$. The covariance operator $\mathcal{C}=\ee(X\otimes X)$ is then compact and self-adjoint, with eigendecomposition $\mathcal{C}=\sum_{i=1}^\infty\lambda_i\phi_i\otimes\phi_i$. Throughout the paper, $\{(\lambda_i,\phi_i)\}$ denotes exclusively the eigensystem of $\mathcal{C}$ in FPCA. The operator $\mathcal{C}$ corresponds to the covariance function $C(s,t)$, whose spectral decomposition is $C(s,t)=\sum_{i=1}^\infty\lambda_i\phi_i(s)\phi_i(t)$. Since $\{\phi_i\}$ forms an orthonormal basis of $L_2([0,1])$, the random function $X$ admits the Karhunen--Lo\`eve expansion $X=\sum_{i=1}^\infty\xi_i\phi_i$, where $\xi_i=\langle X,\phi_i\rangle_{L_2}$. FPCA represents $X$ by the first $\mdd$ coefficients $\xi_i,i=1,...,\mdd$.

Denote the FPCA encoder by $\fpca(X)=(\xi_1,...,\xi_\mdd)^\t$. Then the SDR architecture $f=\net\circ\fpca$, $h\in H_{\mathrm{NN}}$. Define the inverse operator of $\fpca$ by $\fpca^{-1}(\xi)=\sum_{i=1}^\mdd\xi_i\phi_i$, where $\xi=(\xi_1,...,\xi_\mdd)^\t$ contains the  Karhunen--Lo\`eve coefficients of $X$.

\subsubsection{Approximation rates}
The FPCA analysis involves the reconstruction error $\|(\mathcal{I}-\fpca^{-1}\fpca)X\|_{L_2}^2$, where $\mathcal{I}$ is the identity operator. Because $\fpca$ depends on $P_X$, we bound $\ee_X\|(\mathcal{I}-\fpca^{-1}\fpca)X\|_{L_2}^2$ rather than $\sup_{X\in H_X}\|(\mathcal{I}-\fpca^{-1}\fpca)X\|_{L_2}^2$. Below we relate this distributional formulation to the deterministic smoothness conditions used for kernel embedding. A further issue is that the FPCA scores need not lie in a fixed bounded region, whereas the ReLU approximation argument is applied on such a region. The following boundedness condition on the target will allow us to control the contribution from scores outside it.

\begin{assumption}\label{bounded_f}
    For the target nonlinear functional $f:H_X\rightarrow\real^d$, assume $\|f\|_\infty\leq \mathsf{L}$. 
\end{assumption}

For kernel embedding, the network input $\T(L_KX)$ is controlled through the tunable parameter $\gamma$. There is no analogous deterministic bound on $\fpca X$, so the FPCA argument restricts the network input to a bounded region and uses Assumption~\ref{bounded_f} to control the remaining tail error. Specifically, the input is truncated at $\mathsf{R}$ and the output at $\mathsf{L}$. The same boundedness assumption will be used for both encoders in the generalization analysis of Section~\ref{generalization}. Define $\pi_\mathsf{L}(u)=u\cdot\mathbbm{1}(|u|\leq \mathsf{L})$ for $u\in\real$; for vector-valued $u$, apply $\pi_\mathsf{L}$ componentwise.

\begin{theorem}\label{fpca_appro1}
Suppose that Assumptions~\ref{ass_class}, \ref{F_modulus}, \ref{bounded_f} hold. For $X\sim P_X$, let $\fpca$ be the population FPCA encoder. Assume $\ee\|X\|_{L_2}^2<\infty$ and polynomial eigenvalue decay for $\mathcal{C}$: $\lambda_j\leq cj^{-a}$, $a>1$. For a suitable choice $\mdd\asymp\frac{\log M}{\log\log M}$, there exists a neural operator $ f_{\mathrm{NN}}= \net_{\mathrm{NN}}\circ\fpca$, where $\net_{\mathrm{NN}}$ is a finite-dimensional deep ReLU network with $M$ nonzero parameters and depth $L\lesssim \big(\frac{\log M}{\log\log M}\big)^2$, such that
\begin{equation*}
\ee\| \pi_\mathsf{L}\circ f_{\mathrm{NN}}(X)-f(X)\|_{2}^2\lesssim \left(\frac{\log M}{\log\log M}\right)^{-\lambda(a-1)}.
\end{equation*}
\end{theorem}

The polynomial decay condition $\lambda_j\leq cj^{-a}$ is standard in functional data analysis. Let $A_\mathsf{R}=\{\|\fpca X\|_\infty\leq \mathsf{R}\}$, where $\mathsf{R}$ is a tunable input truncation threshold. The approximation error decomposes as
\begin{equation}\label{decom2}
    \begin{split}
\ee\|\pi_\mathsf{L}\circ f_{\mathrm{NN}}(X)-f(X)\|_{2}\leq &\ee\|f(X)-f\{\fpca^{-1}\fpca(X)\}\|_{2}\\
&+\ee\Big[\|f\fpca^{-1}\{\fpca(X)\}-\pi_{\mathsf{L}}h_{\mathrm{NN}}\{\pi_{\mathsf{R}}\fpca(X)\}\|_{2}\Bigm| \mathbbm{1}(A_\mathsf{R})\Big]\\
&+\ee\Big[\|f\fpca^{-1}\{\pi_\mathsf{R}\fpca(X)\}-\pi_{\mathsf{L}} f_{NN}(X)\|_{2}\Bigm| \mathbbm{1}(A_\mathsf{R}^c)\Big],
    \end{split}
\end{equation}
The first term is the reconstruction error, controlled by $(\sum_{j>\mdd}\lambda_j)^{\lambda/2}$ under the $\lambda$-H\"older condition on $f$. The second is the ReLU approximation error for a $\lambda$-H\"older function $f\fpca^{-1}$ on the bounded input region $A_\mathsf{R}$. On $A_\mathsf{R}^c$, where $\fpca X$ may be unbounded, output truncation controls the third term by $2\mathsf{L}\cdot\mathbb{P}(A_\mathsf{R}^c)$. The decomposition explains why the encoder dimension and input truncation threshold must be chosen together with the network size: they govern the reconstruction error, the region on which the network approximates the target, and the probability of leaving that region.

It seems that the rate in Theorem~\ref{fpca_appro1} is irrelevant to the input space smoothness. This is because the FPCA theorem is written in spectral-distribution language while the kernel embedding theorems are written in deterministic function-space language. However, in fact, the exponent $a$ in the eigenvalue decay has already compressed the effects of smoothness. Single eigenvalue decay $\lambda_j\lesssim j^{-a}$ implies $\sum_{j> r}\lambda_j\lesssim r^{-(a-1)}$ and only the latter enters the proof. The following examples connect input regularity to such spectral conditions.
\begin{itemize}
\item \emph{Mean-square Sobolev smoothness of $P_X$.} Assume $X\in H^\alpha([0,1])\subset B_{2,\infty}^\alpha([0,1])$ in mean square, meaning $\ee\|X\|_{H^\alpha}^2\leq B^2$. This is equivalent to
\begin{equation}\label{fpca_ell2}
\ee\sum_{j\geq 1}\frac{|\langle X,\psi_j\rangle_{L_2}|^2}{(1+\nu_j)^{-\alpha}}\leq B^2,
\end{equation}
where, with a slight abuse of notation, $(\nu_i,\psi_i)$ denotes the eigensystem of $-\Delta$, a nonnegative Laplacian-type operator on $[0,1]$ with a specified boundary condition. Then $\sum_{j>\mdd}\lambda_j\leq B^2\mdd^{-2\alpha}$. The parameter $\alpha$ measures input smoothness, corresponding here to $a=2\alpha+1$.

\item \emph{Mean-square FPCA ellipsoid condition.} Assume
\begin{equation}\label{fpca_ell}
\ee\sum_{j\geq1}\frac{|\langle X,\phi_j\rangle_{L_2}|^2}{\lambda_j^s}\leq B^2,\quad s\neq 1,
\end{equation}
where $(\lambda_i,\phi_i)$ are the eigenpairs of the covariance operator associated with $P_X$. Equivalently, $\sum_{j\geq 1}\lambda_j^{1-s}\leq B^2$, giving $\sum_{j>\mdd}\lambda_j\lesssim \lambda_{\mdd+1}^s\lesssim r^{-as}$. Here $a=1/(1-s)$, with $s$ describing input smoothness.
\end{itemize}
Both (\ref{fpca_ell2}) and (\ref{fpca_ell}) are ellipsoid conditions, paralleling (\ref{source1}) and (\ref{source2}) in the kernel embedding analysis. The difference lies in the eigenpairs $(\nu_j,\psi_j)$ or $(\lambda_j,\phi_j)$ of different operators. Thus both encoders exploit smoothness of the input functions: kernel embedding expresses it through deterministic function-space constraints, while FPCA expresses it through mean-square or spectral conditions.

The preceding examples link input regularity to the spectral tail that controls reconstruction. We now strengthen the eigenvalue decay condition on $\mathcal{C}$ and examine how the improved reconstruction accuracy changes the approximation bound.

\begin{theorem}\label{fpca_appro2}
Suppose that Assumptions~\ref{ass_class}, \ref{F_modulus}, \ref{bounded_f} hold. For $X\sim P_X$, let $\fpca$ be the population FPCA encoder. Assume $\ee\|X\|_{L_2}^2<\infty$ and exponential eigenvalue decay for $\mathcal{C}$: $\lambda_j\leq c e^{-bj}$, $b>0$. For a suitable choice $\mdd\asymp(\log M)^{1/2}$, there exists a neural operator $f_{\mathrm{NN}}= \net_{\mathrm{NN}}\circ\fpca$, where $\net_{\mathrm{NN}}$ is a finite-dimensional deep ReLU network with $M$ nonzero parameters and depth $L\lesssim \log M$, such that
\begin{equation*}
\ee\|\pi_\mathsf{L}\circ f_{\mathrm{NN}}(X)-f(X)\|_{2}^2\lesssim\exp\big[-\{2c_b/(\lambda+1)\}^{1/2}\lambda(\log M)^{1/2}\big](\log M)^{1/(\lambda+1)},
\end{equation*}
where $c_b$ is a positive constant depending on $b$.
\end{theorem}

The proof uses an error decomposition analogous to (\ref{decom2}). Exponential eigenvalue decay improves the rate in Theorem~\ref{fpca_appro2}: it decays faster than $(\log M)^{-l}$ but more slowly than $M^{-l}$, which suggests that even when the eigenvalues decay in a extremely fast exponential rate, indicating a smooth enough input space, we still cannot achieve polynomial rates, due to the infinite-dimensionality of functional data.
There are many realistic examples where random functions exhibit exponential eigenvalue decay.
\begin{itemize}
\item \emph{Processes with Gaussian covariance kernels.} If $X(t)$ has a Gaussian covariance kernel, its FPCA eigenvalues satisfy $\lambda_j\lesssim e^{-cj}$ for some $c>0$.
\item \emph{Smoothed random fields.} Let $W$ be white noise on [0,1], interpreted as a generalized random field, and let $X=e^{-\tau A}W$ where $\tau>0$ and $A=-\Delta$ is the Laplacian with homogeneous Dirichlet boundary conditions. The covariance operator of $X$ is $\mathcal{C}=e^{-2\tau A}$. Since the eigenvalues of $A$ are given by $\lambda_j(A)=\pi^2 j^2$, the covariance eigenvalues satisfy $\lambda_j= e^{-2\tau\pi^2 j^2}$. This construction provides a natural stochastic model for heat-diffused or strongly smoothed signals, closely related to Gaussian scale-space representations in image processing \cite{Koen1984} and heat-kernel Gaussian processes for spatial random fields \cite{Niuetal2019}.
\item \emph{Random functions in a Gaussian RKHS.} This parallels the $H_\gamma([0,1])$ assumption in Theorem~\ref{kernel_appro_2}. Let $k_\gamma(s,t)=\sum_{j\geq 1}\nu_j\psi_j(s)\psi_j(t)$ with $\nu_j\lesssim e^{-cj}$, and assume $\ee\|X\|_{H_\gamma}^2\leq B^2$. Then $\sum_{j\geq\mdd}\lambda_j\lesssim e^{-c\mdd}$. Gaussian RKHS regularity thus yields an exponential FPCA tail without requiring explicit knowledge of the FPCA basis. This example also illustrates the connection between FPCA language and kernel embedding language, both imposing smoothness conditions on the input function space.
\item \emph{Finite-rank processes.} Let $X(t)=\sum_{j=1}^R\xi_j\psi_j(t)$ with a fixed integer $R\geq 1$. This imposes an even stronger restriction than exponential decay and describes functions generated by a small number of latent modes, as in low-dimensional dynamical systems or structured experimental curves.
\end{itemize}

\section{Generalization analysis}\label{generalization}
The approximation results in Section 4 describe what the encoder--network architecture can represent at the population level. Now we study the additional errors that arise when the network is learned from discrete noisy observations. We formulate this problem within empirical risk minimization (ERM). Assume the first-stage sample $D=\{X_i,Y_i\}_{i=1}^n$ are i.i.d. samples drawn from the unknown Borel probability distribution $P_{XY}$ on $H_X\times H_Y$, where $H_X$ is the input function space and $H_Y=\real^q$. Assume $\kappa(y,\tilde{y})\leq \kappa_B$ for all $y,\tilde{y}\in H_Y$.

The complete trajectories in $D$ are unavailable. Instead, we observe the second-stage sample $\widehat{D}=\{(t_{ij},X_{ij})_{j=1}^{N_i},Y_i\}_{i=1}^n$, where
\begin{equation*}
X_{ij}=X_i(t_{ij})+\varepsilon_{ij},\quad j=1,...,N_i.
\end{equation*}
Each trajectory $X_i(t)$ is measured at $N_i$ time points. The measurement errors $\varepsilon_{ij}$ are copies of $\varepsilon$, with $\ee(\varepsilon)=0$ and $\vv(\varepsilon)=\sigma_X^2$, and $\{\varepsilon_{ij}\}_{i,j}$ are independent of $X_i$.

Following classical learning theory, our population objective is to learn the oracle nonlinear SDR operator $f_0\in \{\L_2(P_{X})\}^d$ that minimizes the regularized GMDD loss:
\begin{equation*}
\mathcal{E}(f)=-\ee\left([f(X)-\ee\{f(X)\}]^\t[f(\widetilde{X})-\ee\{f(\widetilde{X})\}]\kappa(Y,\widetilde{Y})\right)+\eta\|\mathrm{var}\{f(X)\}-I_d\|_F^2,
\end{equation*}
where $(\widetilde{X},\widetilde{Y})$ is an independent copy of $(X,Y)$.

From now on, we follow \cite{Shi2025} to equip the ReLU network class $H_{\mathrm{NN}}$ a specific strcuture. This class will be used with both encoders. Each $h\in H_{\mathrm{NN}}$ has input dimension $\mdd$, input domain $[-\mathsf{R},\mathsf{R}]^\mdd$, depth $L=\mdd^2+\mdd+1$, and $M$ nonzero parameters. The $j$-th output neuron of $h$ takes the form
\begin{equation}\label{output_relu}
[\net(u)]_j=\sum_{i=1}^{(\mathsf{G}+1)^\mdd}c_{ji}\vartheta\big(\frac{\mathsf{G}}{2\mathsf{R}}(u-b_{i})\big),\quad u\in[-\mathsf{R},\mathsf{R}]^\mdd,
\end{equation} 
where $c_{ji}\in\real,b_i\in\real^\mdd$ satisfy $|c_{ji}|\leq C_{\mathrm{NN}}$ and $\|b_i\|_\infty\leq B_{\mathrm{NN}}$, and $\vartheta:\real^\mdd\rightarrow\real$ is defined by
\begin{equation*}
    \vartheta(u)=\sigma\Big( \min\{\min_{k\neq j}(1+u_k-u_j),\min_k (1+u_k),\min_k (1-u_k)\}\Big).
\end{equation*}
In (\ref{output_relu}), the integer $\mathsf{G}$ specifies the mesh resolution for approximating H\"older continuous functions on $[-\mathsf{R},\mathsf{R}]^\mdd$. It determines the number $(\mathsf{G}+1)^\mdd$ of local interpolation units in $H_{\mathrm{NN}}$. Increasing $\mathsf{G}$ improves approximation by refining the partition, while increasing the number of nonzero parameters, which satisfies $M\asymp \mdd^4(\mathsf{G}+1)^\mdd$; see Lemma 14 of \cite{Shi2025}.

\subsection{Kernel embedding}
For kernel embedding, define the hypothesis class $\H_{\mdd,M}$ by $\H_{\mdd,M}=$\{$\net\circ \T$: $h\in H_{\mathrm{NN}}$, $\T$ is the projection operator induced by $L_{K_0}^\mu$\}.

To better demonstrate the decompositino of generalization error, in the following we define several types of errors corresponding to different stages. The empirical generalization error using the first-stage sample $D$ is defined as 
\begin{equation*}
\mathcal{E}_D(f)=\mathrm{GMDD}_n\{f(X_i),Y_i\}+\eta\|\mathrm{var}_n\{f(X_i)\}-I_d\|_F^2.
\end{equation*}
If we employ the hypothesis space $\H_{\mdd,M}$ along with the kernel embedding step, the first-stage empirical generalization error can be formulated as
\begin{equation*}
\mathcal{E}_D(H\circ L_K)=\mathrm{GMDD}_n\{H( L_K X_i),Y_i\}+\eta\|\mathrm{var}_n\{H( L_K X_i)\}-I_d\|_F^2,
\end{equation*}
where $H\in\H_{\mdd,M}$. Define
\begin{equation*}
H_D=\mathop{\arg\min}_{H\in\H_{\mdd,M}}\mathcal{E}_D(H\circ L_K),
\end{equation*}
then the first-stage empirical target operator is $f_D=H_D\circ L_K$.

However, as previously noted, we can only acquire the second-stage sample $\widehat{D}$. Therefore, the integral $L_KX_i$ must be approximated by discrete observations. To maintain generality, we assume the time points $\{t_{ij}\}$ are i.i.d. sampled from the uniform distribution on $[0,1]$ and approximate $L_KX_i$ by standard Monte Carlo method:
\begin{equation}\label{quadra}
\widehat{L_KX_i}=\sum_{j=1}^{N_i}\frac{1}{N_i}K(\cdot,t_{ij})X_{ij},\quad i=1,...,n.
\end{equation}
Approximating (\ref{quadra}) with $N_i$ sample points gives an error rate of $O(1/\sqrt{N_i})$. Alternating sampling methods such as Quasi Monte-Carlo and Trapezoidal rules can lead to improved convergence rates. In \cite{Shi2025}, $L_KX_i$ was approximated by $\sum_{j=1}^{N_i}\theta_{ij}K(\cdot,t_{ij})$ where $t_{ij}$ and $\theta_{ij}$ are chosen to minimize the approximation error as much as possible. This leads to negligible quadrature approximation error but is unrealistic for real-world functional data.

The second-stage empirical generalization error is
\begin{equation*}
\mathcal{E}_{\widehat{D}}(H\circ L_K)=\mathrm{GMDD}_n\{H(\widehat{L_KX_i}),Y_i\}+\eta\|\mathrm{var}_n\{H(\widehat{L_KX_i})\}-I_d\|_F^2.
\end{equation*}
Note that $L_K$ requires no estimation. If we denote 
\begin{equation*}
H_{\widehat{D}}=\mathop{\arg\min}_{H\in\H_{\mdd,M}}\mathcal{E}_{\widehat{D}}(H\circ L_K),
\end{equation*}
then the second-stage empirical target function has the form $f_{\widehat{D}}=H_{\widehat{D}}\circ L_K$.

Finally, we assume the orcle SDR oeprator $f_0$ is bounded by $\mathsf{L}$, and use the truncated empirical target functional $\pi_\mathsf{L}\circ f_{\widehat{D}}$ as the final estimator.

\begin{equation*}
\loss(\pi_{\mathsf{L}}f_{\widehat{D}})-\loss(f_0)\leq I_1+I_2+I_3,
\end{equation*}
where
\begin{equation*}
    \begin{split}
I_1=&\{\loss(\pi_\mathsf{L}f_{\widehat{D}})-\loss((f_0))\}-2\{\loss_{D}(\pi_\mathsf{L}f_{\widehat{D}})-\loss_D(f_0)\},\quad I_2=2\{\loss_D(\pi_\mathsf{L}f_D)-\loss_D(f_0)\},\\
I_3=&4\mathsf{L}\kappa_B\sup_{H\in\H_{\mdd,M}}\frac{1}{n}\sum_{i=1}^n\|H(L_KX_i)-H(\widehat{L_KX_i})\|_2.
    \end{split}
\end{equation*}
In the decomposition, $I_1$ is the typical empirical process error; $I_2$ can be bounded by the approximation error established in Section~\ref{kernel_app}; and $I_3$ contains the quadrature approximation error and noise contamination when calculating $L_KX_i$. Since $\loss$ is the GMDD loss and $\loss_{D}$ is a U-statistic of order 4, Hoeffding decomposition is applied when analyzing $I_1,I_2,I_3$.

\begin{theorem}\label{excess_kernel1}
Suppose Assumptions~\ref{ass_class}, \ref{F_modulus}, \ref{bounded_f} hold, and $\|x\|_{B_{2,\infty}^\alpha}\leq 1$ for every $x\in H_X$. Assume equal second-stage sample sizes, $N_1=...=N_n=N$. Choose the number of nonzero parameters in $f_{\widehat{D}}$ as $M\asymp n^\rho$, with $0<\rho<\frac{1}{2}$, and choose the second-stage sample size $N$ to satisfy
\begin{equation*}
N\gtrsim  n^{2\rho}(\log n)^{2(\alpha\lambda+1/\tilde{c}_1)}(\log\log n)^{-8\alpha\lambda},
\end{equation*}
where $\tilde{c}_1$ is a positive constant. Then
\begin{equation*}
\ee\{ \mathcal{E}(\pi_\mathsf{L}\circ f_{\widehat{D}})-\mathcal{E}(f_0)\}\lesssim (\log n)^{-2\alpha\lambda}(\log\log n)^{4\alpha\lambda}.
\end{equation*}
\end{theorem}

The parameter choices describe a tradeoff between network complexity and observation frequency. In particular, $\rho$ controls the required second-stage sample size. Sparse regimes in functional data analysis are often characterized by observation counts below $O(n^{1/4})$ \cite{ZhangWang2016,ZhouWeiYao2025}. Theorem~\ref{excess_kernel1} nevertheless requires sufficiently many observations to support the chosen network complexity. Taking $\rho=\frac{1}{8}$ illustrates the practical limitation: at $n=10000$, the prescribed parameter order corresponds to approximately $3$ parameters. Reducing observation requirements through a very small exponent can therefore leave too little network capacity at realistic sample sizes.

As we mentioned before, \cite{Shi2025} assumed an optimal quadrature scheme with noiseless samples when approximating $L_KX_i$, which results in a second-stage sample size $N=O(\log n)$. Since this assumption is unrealistic, for generality we consider uniformly sampled i.i.d. time points and noise contanimation for the second-stage sample, and adopt the standard Monte Carlo method. Thus $N=(n^{2\rho})$ is required to achieve satisfatory accuracy.

The generalization error is the summation of $I_1$ (too large if network too big), $I_2$ (too large if network too small) and $I_3$ (too large if $N$ too small). \cite{Shi2025} chose the number of the network parameters $M$ to be linear in $n$ (up to logarithmic term). In Theorem~\ref{excess_kernel1}, we reduce the number of network parameters to $m^\rho$ with $0<\rho<1$ to keep the final rate same as theirs. The reason is that their quadrature error is so cheap (controlled with only polylogarithmic $N$) that they need not even control it. So they increase $M$ as large as possible until $I_1$ balances $I_2$. But our quadrature approximation error is expensive. If $N$ is limited (we want it at most linear in $n$), to control this term we must deliberately reduce $M$ because a smaller $M$ makes the learned neural operator less sensitive to noisy and discretized kernel embeddings. That is, we are balancing $I_2$ and $I_3$ instead.

The resulting bound is polynomial in $\log n$, consistent with the Besov approximation result in Section 4. We next impose stronger smoothness on the input functions and use the corresponding approximation bound to obtain a faster generalization rate.

\begin{theorem}\label{excess_kernel2}
Suppose Assumptions~\ref{ass_class}, \ref{F_modulus}, and \ref{bounded_f} hold. Assume that $H_X$ is contained in the unit ball of $H_\gamma([0,1])$ and $N_1=...=N_n=N$. Choose $M\asymp n^\rho$, with $0<\rho<\frac{1}{2}$, as the number of nonzero parameters in $f_{\widehat{D}}$, and choose $N$ to satisfy
\begin{equation*}
N\gtrsim n^{2\rho}e^{c_\rho(\log n)^{1/2}},
\end{equation*}
where $c_\rho$ is a positive constant. Then
\begin{equation*}
\ee\{ \loss(\pi_\mathsf{L}\circ f_{\widehat{D}})-\loss(f_0)\}\lesssim \exp\left\{ -2\sqrt{1/3}\lambda\rho^{1/2}(\log n)^{1/2}\right\}\log n,
\end{equation*}

\end{theorem}

The stronger input smoothness improves the bound in Theorem~\ref{excess_kernel2} relative to the Besov case. The resulting rate remains slower than $n^{-l}$ for every $l>0$. As in the approximation analysis, this is the bound obtained while retaining general regularity of the target operator on an infinite-dimensional input class.

\subsection{Discrete FPCA}
For FPCA, evaluating the encoded coordinates also requires estimating the encoder itself. Given discrete noisy measurements $\{(t_{ij},X_{ij})\}$, we estimate the covariance function $\hat{C}(s,t)$ by local linear smoothing \cite{ZhangWang2016,ZhouYaoZhang2023,ZhouWeiYao2025}:
\begin{equation*}
    \begin{split}
(\hat{\beta}_0,\hat{\beta}_1,\hat{\beta}_2)=&\mathop{\arg\min}_{\beta_0,\beta_1,\beta_2}\sum_{i=1}^nv_i\sum_{1\leq l_1\neq l_2\leq N_i}\left\{X_{il_1}X_{il_2}-\beta_0-\beta_1(t_{il_1}-s)-\beta_2(t_{il_2}-t)\right\}^2\\
& \quad \times\frac{1}{h}\kk\left(\frac{t_{il_1}-s}{h}\right)\kk\left(\frac{t_{il_2}-t}{h}\right),
    \end{split}
\end{equation*} 
where $v_i=\{nN_i(N_i-1)\}^{-1}$, $\kk$ is a symmetric Lipschitz continuous density kernel on $[-1,1]$, and $h$ is the bandwidth. Then $\hat{C}(s,t)=\hat{\beta}_0$. Its spectral decomposition is
$\hat{C}(s,t)=\sum_{k=1}^\infty\hat{\lambda}_k\hat{\phi}_k(s)\hat{\phi}_k(t)$, where $\hat{\lambda}_k$ and $\hat{\phi}_k$ are the estimated eigenvalues and eigenfunctions, respectively. We reserve $\kk$ for the smoothing kernel used in discrete FPCA.

Once the basis has been estimated, the principal component scores of $X_i$, defined by $\hat{\xi}_k=\int_0^1X_i(t)\hat{\phi}_k(t)dt$, are evaluated from the discrete measurements by Monte Carlo integration. We call the resulting encoder discrete FPCA and denote it by $\widehat{\fpca}$. Its presmoothing step pools observations across curves to estimate a smooth covariance surface, followed by a truncated expansion in the estimated eigendirections. Thus, the two encoders use presmoothing at different levels: kernel embedding smooths individual inputs through a fixed operator, whereas discrete FPCA learns a common representation from the pooled covariance. The latter introduces basis-estimation error in addition to the numerical integration error. 

As in the kernel-embedding analysis, we distinguish an ideal learner using the population encoder and complete curves from the learner constructed with the estimated encoder and discrete observations. Define
\begin{equation*}
    \begin{split}
h_D=& \mathop{\arg\min}_{h\in H_{\mathrm{NN}}}\loss_D(h\circ\fpca)=\mathop{\arg\min}_{h\in H_{\mathrm{NN}}}\mathrm{GMDD}_n\big(h\circ\fpca(X_i),Y_i\big)+\eta\|\mathrm{var}_n\{h\circ\fpca(X_i)\}-I_d\|_F^2,\\
h_{\widehat{D}}=& \mathop{\arg\min}_{h\in H_{\mathrm{NN}}}\loss_{\widehat{D}}(h\circ\widehat{\fpca})=\mathop{\arg\min}_{h\in H_{\mathrm{NN}}}\mathrm{GMDD}_n\big(h\circ\widehat{\fpca}(X_i),Y_i\big)+\eta\|\mathrm{var}_n\{h\circ\widehat{\fpca}(X_i)\}-I_d\|_F^2,
    \end{split}
\end{equation*}
where $\widehat{\fpca}$ is estimated from the same sample $\{X_i\}_{i=1}^n$ used to train the network, and $H_{\mathrm{NN}}$ has the structure in (\ref{output_relu}). In particular, encoder estimation and network training are not separated by sample splitting. The final estimator is $\pi_\mathsf{L}\circ f_{\widehat{D}}=\pi_\mathsf{L}\circ h_{\widehat{D}}\circ\widehat{\fpca}$.

Accordingly, the excess risk decomposition is:
\begin{equation*}
\loss(\pi_{\mathsf{L}}f_{\widehat{D}})-\loss(f_0)\leq I_1+I_2+I_3+I_4
\end{equation*}
where
\begin{equation*}
    \begin{split}
I_1=&\{\loss(\pi_\mathsf{L}h_{\widehat{D}}\fpca)-\loss((f_0))\}-2\{\loss_{D}(\pi_\mathsf{L}h_{\widehat{D}}\fpca)-\loss_D(f_0)\},\quad I_2=2\{\loss_D(\pi_\mathsf{L}h_{\widehat{D}}\fpca)-\loss_D(f_0)\},\\
I_3=&4\mathsf{L}\kappa_B\sup_{h\in H_{\mathrm{NN}}}\frac{1}{n}\sum_{i=1}^n\|h(\fpca X_i)-h(\widehat{\fpca}X_i)\|_2,\quad I_4=|\loss(\pi_\mathsf{L}h_{\widehat{D}}\widehat{\fpca})-\loss(\pi_\mathsf{L} h_{\widehat{D}}\fpca)|.
    \end{split}
\end{equation*}

In the decomposition, $I_4$ can be controlled by $I_3$; $I_1$ is the empirical process error; $I_2$ is the approximation error; and $I_3$ is the encoder error that is controlled by
\begin{equation*}
    \ee I_3\lesssim  M^{1+1/\mdd}\Big\{\Big(\sum_{j=1}^\mdd\ee\|\phi_j-\hat{\phi}_j\|_{L_2}^2\}\Big)^{1/2}+\frac{1}{n}\sum_{i=1}^n\frac{1}{\sqrt{N_i}}\Big\},
\end{equation*}
The first term accounts for discrete FPCA estimation and the second for numerical integration; both include the effect of measurement noise. The additional contribution $\sum_{j=1}^\mdd\ee\|\phi_j-\hat{\phi}_j\|_{L_2}^2$ arises because the FPCA basis is estimated from data. For kernel embedding, $L_K$ and $\T$ are fixed in advance. The dependence on $\mdd$ also matters: improving the population approximation by retaining more coordinates requires controlling the accumulated error in more estimated eigenfunctions.

An explicit excess risk bound therefore requires nonasymptotic control of $\ee\|\hat{\phi}_j-\phi_j\|_{L_2}^2$. However, to the best of our knowledge, there is no such nonasymptotic results with discretely observed samples. Existing literature only focuses on establishing asymptotic bounds for discretely observed functional data \cite{ZhouYaoZhang2023,ZhangWang2016}, or deriving much easier nonasymptotic bounds for fully observed one. Recentlty, \cite{ZhouWeiYao2025} established probably the state-of-the-art convergent result for $\|\hat{\phi}_j-\phi_j\|_{L_2}^2$ with diverging index $j$ by a newly developed device called double truncation. Their setting fits ours well since our network input dimension (or encoder output dimension) $\mdd$ depends on the sample size $n$ and needs to be tuned. Unfortunately their results are also asymptotic. So before obtaining the generalization error bound, based on results of \cite{ZhouWeiYao2025} we first prove a nonasymptotic bound for $\ee\|\hat{\phi}_j-\phi_j\|_{L_2}^2$, during whose proof the double truncation technique was retained.

\begin{assumption}\label{addition}
    We make the following assumptions.
    \begin{itemize}
    \item[(I)] $\kk$ is a symmetric Lipschitz probability density supported on $[-1,1]$ with $0\leq \kk\leq \kk_\infty$. Also, $\kk$ has positive mass in two separated subintervals of $[0,1]$; for example, $\int_0^{1/4}\kk(t)dt>0$ and $\int_{1/4}^{1/2}\kk(t)dt>0$.
    \item[(II)] The design points $t_{ij}$ are i.i.d. on $[0,1]$, independent of $X$ and $\varepsilon_{ij}$, with density $v$ satisfying $0<m_f\leq v(t)\leq M_f<\infty$ and $\|v'\|_\infty+\|v''\|_\infty<\infty$.
    \item[(III)] (i) The covariance kernel $C$ is twice continuously differentiable on $[0,1]^2$, with bounded derivatives up to order $2$; (ii) For some $a>1$, $c j^{-a}\leq\lambda_j\leq Cj^{-a}$, $\lambda_j-\lambda_{j+1}\geq c j^{-a-1}$; (iii) For some $\mathfrak{c}>0$, $\|\phi_j\|_\infty\leq C_\phi$, $\|\phi_j'\|_\infty\leq C_\phi j^{\mathfrak{c}/2}$, $\|\phi_j''\|_\infty\leq C_\phi j^\mathfrak{c}$; (iv) $\ee \xi_{ik}^4\leq c\lambda_k^2$.
    \item[(IV)] For some $\omega>3$, $\ee\sup_{t\in [0,1]}|X_i(t)|^{2\omega}<\infty$, $\ee |\varepsilon_{ij}|^{2\omega}<\infty$.
    \item[(V)] For the bandwidth $0<h\leq 1/4$ and a mesh exponent $\epsilon>0$, assume $\Delta_n^\sharp\leq c_*$, $n^{-\epsilon}h^{-3}\leq C_*\Delta_n^\sharp$, $\mathfrak{t}_{n,\omega}(h)\leq C_*$.
    \item[(VI)]  $h^4j^{2a+2\mathfrak{c}}\leq B_1$, $hj^a\log n\leq B_2$, where $B_1,B_2$ are fixed constants. 
    \end{itemize}
\end{assumption}

Most conditions in Assumption~\ref{addition} follow \cite{ZhouWeiYao2025} with two strengthed here. First, the positive mass in two separated subintervals in (I) is made to derive deterministic lower bounds and is satisfied by all common kernels. Second, the $C^2$ condition in (II) is stronger than the random-design assumption in \cite{ZhouWeiYao2025}, but it is needed for the finite-sample Taylor bounds. Detailed clarification of these conditions can be found in Appendix F.1 in the Supplement, along with definitions of notations used in the assumption and below. 
For $j\leq m$, with $m$ a diverging integer, define the perturbative event $\Omega_m(n,N,h)=\{\|\widehat{C}-C\|_{\mathrm{HS}}\leq \eta_m/2\}$, where $\eta_m=\min_{k\neq m}|\lambda_k-\lambda_m|$. Propositions~\ref{eigenfunc} is established within $\Omega_m(n,N,h)$.


\begin{proposition}\label{eigenfunc}
Under Assumption~\ref{addition}, for every $A>0$, there exists $C_A<\infty$ such that
\begin{equation*}
\ee\|\hat\phi_j^{\mathrm{tr}}-\phi_j\|_{L_2}^2\leq
C_A\left\{
M_{n,h}(j)
+j^{2a+2}R_{n,h}(A)
\right\},
\end{equation*}
where
\begin{equation*}
    \begin{split}
M_{n,h}(j)=&
\frac{j^2}{n}\left(1+\frac{j^a}{\bar{N}_2}+\frac{j^{2a}}{\bar{N}_2^2}\right)
+
\frac{j^a}{n\bar{N}_2 h}\left(1+\frac{j^a}{\bar{N}_2}\right)
+h^4j^{2\mathfrak{c}+2},\\
R_{n,h}(A)=&(h+\Gamma_n)\left\{h^4+\Delta_{n,\alpha}^{(2),0}\right\}+h^{-4}n^{-A}.
    \end{split}
\end{equation*}
\end{proposition}

The superscript ``tr'' is introduced only for technical reasons in the proof. More precisely, let \(\mathcal D_n\) denote the exceptional event on which the local-linear normalizing quantities are not uniformly well controlled, for instance when some denominator in the local-linear covariance estimator becomes too small. In the proof, we work with the trimmed covariance estimator $\widehat C^{\mathrm{tr}}=\widehat C\,\mathbbm{1}(\mathcal D_n^\mathrm{c})$, and define \(\hat\phi_j^{\mathrm{tr}}\) as the corresponding empirical eigenfunction. On the high-probability event \(\mathcal D_n^c\), the trimmed and untrimmed covariance estimators coincide, and hence \(\hat\phi_j^{\mathrm{tr}}=\hat\phi_j\). On \(\mathcal D_n\), the estimator may be defined arbitrarily in a measurable way, since this event has probability of order \(O(n^{-A})\) for any prescribed fixed \(A>0\). Therefore, the trimming does not modify the practical discrete FPCA procedure: in implementation, one simply uses the original local-linear covariance estimator \(\widehat C\) and its eigenfunctions \(\hat\phi_j\). The trimming is only a proof device that prevents possible instability of inverse local-linear coefficients on a negligible event and allows us to establish the nonasymptotic expectation bound. So from now on, in the nonparametric estimation of FPCA, $\hat{\phi}_j$ is understood as $\hat\phi_j^{\mathrm{tr}}$.


We now return to the learning problem. Summing the eigenfunction bounds over the retained components controls the basis-estimation contribution to $I_3$. Combining this with the quadrature, empirical-process, and approximation bounds gives the following result.

\begin{theorem}\label{excess_pca1}
Suppose Assumptions~\ref{ass_class}, \ref{F_modulus}, \ref{bounded_f}, and \ref{addition} hold, and $N_1=...=N_n=N$. For $\rho$ satisfying $0<\rho\leq \min(\frac{1}{2},\frac{4\omega-8}{3\omega+8})$, choose the input dimension $\mdd$, the number $M$ of nonzero parameters in $h_{\widehat{D}}$, the second-stage sample size $N$, and the local linear bandwidth $h$ as follows:
\begin{equation*}
    \mdd\asymp \frac{\log M}{\log\log M}, \quad M\asymp\frac{n^\rho}{(\log n)^{B_0}}\mdd^{-\lambda(a-1)-\frac{1}{\lambda+1}}, \quad N\asymp n^{2\rho}\mdd^{-2}, \quad h\asymp n^{-\rho/2}\mdd^{-\mathfrak{c}/2},
\end{equation*}
where $B_0$ is a positive constant. Then
\begin{equation*}
\ee\{ \loss(\pi_\mathsf{L}\circ f_{\widehat{D}})-\loss(f_0)\}\lesssim (\log n)^{-\lambda(a-1)}(\log\log n)^{-\lambda(a-1)}.
\end{equation*}
\end{theorem}

The bottleneck is the approximation error. So in Theorem~\ref{excess_pca1}, we choose $M,h,N$ to make the discrete FPCA encoder error term bounded by the approximation error $\mdd^{-\lambda(a-1)}$ which also controls the first-stage empirical process term and the quadrature error term. The exponent $a$ describes covariance eigenvalue decay and hence the concentration of variation in the leading components. Its connection to input smoothness, discussed in Section 4, explains how regularity of the functional predictor enters the final learning rate.

The approximation results in Section 4 suggest a faster rate under exponential eigenvalue decay. We use the following modification of Assumption~\ref{addition}.
\begin{assumption}\label{exp_decay}
    Retain Assumption~\ref{addition} except (III)(ii). Additionally assume that
    \begin{itemize}
    \item[(I)] $c e^{-bj}\leq\lambda_j\leq C e^{-bj}$ and $\min_{k\neq j}|\lambda_j-\lambda_k|\geq c e^{-bj}$ for some constant $b>0$.
    \item[(II)] $h^2+h^8j^{4\mathfrak{c}}\leq C\lambda_j^2$, where $\mathfrak{c}$ is from Assumption~\ref{addition} (III)(iii).
    \end{itemize}
\end{assumption}

Under Assumption~\ref{exp_decay}, we can derive a nonasymptotic bound for $\ee\|\hat{\phi}_j-\phi_j\|_{L_2}^2$ analogously to Proposition~\ref{eigenfunc}. Due to space limit, we leave it to Appendix A.3 where more explanations about the assumption and the bound are presented. 
Using this eigenfunction bound together with the exponential-decay approximation result yields the following generalization bound.

\begin{theorem}\label{excess_pca2}
Suppose Assumptions~\ref{ass_class}, \ref{F_modulus}, \ref{bounded_f}, and \ref{exp_decay} hold, and $N_1=...=N_n=N$. For $\rho$ satisfying $0<\rho\leq \min(\frac{1}{2},\frac{4\omega-8}{3\omega+8})$, choose the input dimension $r\asymp\sqrt{\log M}$. Set the number $M$ of nonzero parameters in $h_{\widehat{D}}$, the second-stage sample size $N$, and the local linear bandwidth $h$ as follows:
\begin{equation*}
  \begin{split}
    M\asymp n^\rho, \quad N\asymp n^{2\rho}\exp(B_N\sqrt{\log n}),\quad h\asymp n^{-\rho/2}\exp(-B_h\sqrt{\log n})r^{-\mathfrak{c}},
  \end{split}
\end{equation*}
Then
\begin{equation*}
\ee \{\loss(\pi_\mathsf{L}\circ f_{\widehat{D}})-\loss(f_0)\}\lesssim \exp(-c_*\sqrt{\log n}),
\end{equation*}
where $B_N,B_h,c_*$ are positive constants. 
\end{theorem}

The convergence rate is improved to $\exp(-c_*\sqrt{\log n})$ which is faster than any $(\log n)^{-l}$ but still inferior to $n^{-l}$ for any $l>0$. The choice of encoder output dimension also respects the accuracy available for the empirical eigensystem. Zhou, Wei and Yao (2025) noted that exponential eigenvalue decay restricts the number of accurately estimable eigenfunctions to order $\log n$. Our choice of input dimension $\mdd$ at order $\sqrt{\log n}$ lies within that scale, so the faster approximation is used without requiring an encoder dimension beyond this range.

\section{Simulation studies}
\subsection{Dimension determination}\label{determine}
In practice, the dimension $d$ of the central class is unknown and must be estimated. We determine $d$ based on the decay of eigenvalues \cite{Ferre1998,Li1991,Schott1994}. Take GMDD-Net for an example. First set the output dimension of the ReLU network as $m$ that can be suitably large. After obtaining the estimated encoder $\widehat{Q}$ and the trained network $\hat{h}$, calculate an estimator of the GMDD matrix by
\begin{equation*}
\widehat{\Delta}=\binom{n}{2}^{-1}\sum_{j<k}\Big\{\hat{h}\circ\widehat{Q}(X_j)-\frac{1}{n}\sum_{i=1}^n\hat{h}\circ\widehat{Q}(X_i) \Big\}\Big\{\hat{h}\circ\widehat{Q}(X_k)-\frac{1}{n}\sum_{i=1}^n\hat{h}\circ\widehat{Q}(X_i) \Big\}^\t\kappa(Y_j,Y_k).
\end{equation*}
Let $\{\hat{\lambda}_i\}_{i=1}^m$ denote the eigenvalues of $\widehat{\Delta}$. We select $\hat{d}$ as the smallest integer $k$ for which the cumulative eigenvalue ratio satisfies
\begin{equation*}
\sum_{j=1}^k\hat{\lambda}_j\Big/ \sum_{j=1}^m\hat{\lambda}_j\geq \alpha,
\end{equation*}
where we set $\alpha$ to 0.95. When the encoder is independent of the sample, replace $\widehat{Q}$ by $Q$. If FCCov-Net is used, compute the FCCov matrix instead. In the following numerical studies, we always first apply the dimension determination method to obtain $\hat{d}$ and then train the SDR operator with this output dimension.
\subsection{Euclidean response}\label{simu1}
We first consider Euclidean responses. Chosen encoders include kernel embedding and discrete FPCA. We also include the Fourier basis expansion as another encoder for comparison. The SDR network structures include GMDD-Net for Euclidean responses and FCCov-Net for metric-space responses. We select f-GSIR and f-GSAVE \cite{LiSong2017} as benchmarks, which are the only two existing methods dealing with functional nonlinear SDR with discrete observations, to our best knowledge. We consider the following four models:
\begin{equation*}
    \begin{split}
\text{I:}\quad &Y=\langle b_1, X\rangle+\langle b_2, X\rangle+\epsilon, \quad \epsilon\sim N(0,0.1^2);\\
\text{II:}\quad&Y=\Big(\frac{\langle b_1,X\rangle}{1+\exp(\langle b_2,X\rangle)},0.08\langle X,X\rangle\Big)^\t+\epsilon, \quad \epsilon\sim N(0,0.1^2 I_2);\\
\text{III:}\quad&Y=\Big((1+\langle b_1,X\rangle)^2+\epsilon_1,\sin(\frac{\pi}{10}\langle b_2,X\rangle)+\langle b_2, X\rangle^2+\epsilon_2,\epsilon_3,...,\epsilon_{10}\Big)^\t,\epsilon\sim N(0,0.5^2 I_{10});\\
\text{IV:}\quad&Y=\mathbbm{1}\Big(\log\frac{\langle X,X\rangle}{|\langle b_1,X^2\rangle|}+\epsilon>0\Big), \quad\epsilon\sim N(0,0.5^2). 
    \end{split}  
\end{equation*}
In Model IV, $\langle b_1,X^2\rangle$ denotes the $L_2$ inner product. The functions $b_1,b_2$ and the remaining inner products depend on the mechanism used to generate $X$. We consider three mechanisms.
\begin{itemize}
\item[1.] We generate $X(t)$ in the RKHS induced by the Brownian kernel $K(s,t)=\min(s,t)$. Its eigenvalues and eigenfunctions are $\lambda_k=\{(k-0.5)\pi\}^{-2}$ and $e_k(t)=\sqrt{2}\sin\{(k-0.5)\pi t\}$. Set
\begin{equation*}
X(t)=\sum_{k=1}^{100}a_k\sqrt{\lambda_k}Z_ke_k(t),
\end{equation*}
where $Z_k\sim N(0,1)$ and $a_1=1,a_2=3,a_3=...=a_{100}=1$. We set $b_1=e_1,b_2=e_2$ and use the RKHS inner product throughout, except for $\langle b_1,X^2\rangle$.
\item[2.] We generate $X(t)$ in the RKHS induced by the Gaussian kernel $K(s,t)=e^{-7(s-t)^2}$, taking $X$ to be $\sum_{k=1}^5a_kK(\cdot,t_k)$, where $a_1,...,a_5$ are sampled independently from $N(0,1)$ and $t_1,...,t_5$ independently from $U[0,1]$. We choose $b_1(t)=K(t,0.1)$ and $b_2(t)=K(t,0.5)$, and use the RKHS inner product throughout, except for $\langle b_1,X^2\rangle$.
\item[3.] Define $\phi_1(t)=1$, $\phi_k(t)=\sqrt{2}\cos\{(k-1)\pi t\},k\geq 2$. Let
\begin{equation*}
    X(t)=\sum_{k=1}^{50}c_k\phi_k(t),\quad c_k\sim N(0,k^{-2}).
\end{equation*} 
Set $b_1=\text{norm}(\sum_{k=1}^{10}k^{-2}\phi_{2k-1})$ and $b_2=\text{norm}(\sum_{k=1}^{10}k^{-2}\phi_{2k})$, where $\text{norm}(\cdot)$ denotes normalization. All inner products in this mechanism are $L_2$ inner products.
\end{itemize}

Mechanisms 1 and 2 are also used by \cite{LiSong2017}, while Mechanism 3 is common in functional data analysis \cite{LinYao2019,Shi2025}. To simulate the randomly discretely observed functions and to save runtime, we first construct a fixed grid of 101 equally spaced points on $[0,1]$. For each curve, we independently sample 30 grid points uniformly without replacement as its observation times. When implementing the experiments, the layer widths of the ReLU network is set as $10,20,40,20,\hat{d}$, i.e., the output dimension of encoders is 10 and that of the network is the estimated $\hat{d}$. The optimizer is set to ``Adam'' with default parameters, and the learning rate is explictly set as $0.1^3$. The batch size and number of epochs are both configured to be 100. All experiments below are conducted based on such settings unless explicitly stated. For each combination of models and $X$ generation mechanisms, the training sample size $n=1000,2000,3000$. After training, an independent test sample of size 1000 is generated for evaluation. This procedure is repeated for 100 times. 

For each scenario, the dimension determination method is applied on an independent training sample and a test sample both of size 1000, before the experiment. And the estimated $\hat{d}$ for model I-IV are respectively 2,2,2,1 (see Appendix B.1 in the Supplement for details). Note that for model I, if we set $\alpha=0.9$, then all three encoders with GMDD-Net select $\hat{d}=1$. So both $\hat{d}=1$ and 2 are reasonable and we stick to the result of our dimension determination method and set $\hat{d}$ in model I as 2. The means and standard deviations of distance correlations from 100 replicates of each scenario are listed in Table~\ref{table1}. The best mean performance in each row is highlighted in \textbf{bold}.

\begin{table}
\centering
\caption{Mean (standard deviation) of distance correlations over 100 repetitions for three encoders combined with \textbf{GMDD-Net} and for the nested RKHS methods.}
\label{table1}
\begin{tabular}{@{}llcccccc@{}}
\toprule
Model & $X$ generator & $n$ & \shortstack{Fourier\\Expansion} & \shortstack{Kernel\\Embedding} & \shortstack{Discrete\\FPCA} & f-GSIR & f-GSAVE \\
\midrule
\multirow[t]{9}{*}{I} & \multirow[t]{3}{*}{GRB} & 1000 &  0.884(0.019) & 0.966(0.017) & \textbf{0.969}(0.016) & 0.917(0.006) & 0.909(0.009) \\
 &  & 2000 & 0.887(0.019) & 0.962(0.019) & \textbf{0.969}(0.016) & 0.918(0.006) & 0.910(0.008) \\
 &  & 3000 & 0.896(0.015) & \textbf{0.965}(0.019) & 0.960(0.018) & 0.918(0.005) & 0.911(0.008) \\
\cmidrule(l){2-8}
 & \multirow[t]{3}{*}{BMC} & 1000 & 0.962(0.015) & 0.975(0.014) & \textbf{0.976}(0.014) & 0.925(0.005) & 0.909(0.008) \\
 &  & 2000 & 0.953(0.017) & 0.966(0.016) & \textbf{0.970}(0.015) & 0.925(0.004) & 0.912(0.005) \\
 &  & 3000 & 0.952(0.018) & \textbf{0.963}(0.018) & \textbf{0.963}(0.017) & 0.925(0.004) & 0.909(0.007) \\
\cmidrule(l){2-8}
 & \multirow[t]{3}{*}{Cosine} & 1000 & 0.672(0.020) & \textbf{0.951}(0.018) & 0.947(0.019) & 0.815(0.012) & 0.817(0.014) \\
 &  & 2000 & 0.676(0.024) & \textbf{0.947}(0.018) & 0.944(0.018) & 0.815(0.008) & 0.820(0.009) \\
 &  & 3000 & 0.677(0.022) & \textbf{0.948}(0.017) & 0.941(0.018) & 0.814(0.010) & 0.819(0.012) \\
\midrule
\multirow[t]{9}{*}{II} & \multirow[t]{3}{*}{GRB} & 1000 & 0.851(0.034) & \textbf{0.953}(0.019) & 0.929(0.037) & 0.866(0.008) & 0.916(0.013) \\
 &  & 2000 & 0.871(0.013) & \textbf{0.958}(0.013) & 0.955(0.012) & 0.866(0.007) & 0.918(0.008) \\
 &  & 3000 & 0.876(0.012) & \textbf{0.960}(0.011) & 0.958(0.012) & 0.868(0.006) & 0.920(0.009) \\
\cmidrule(l){2-8}
 & \multirow[t]{3}{*}{BMC} & 1000 & 0.702(0.059) & \textbf{0.779}(0.016) & 0.752(0.049) & 0.679(0.016) & 0.498(0.020) \\
 &  & 2000 & 0.725(0.024) & \textbf{0.780}(0.018) & 0.767(0.017) & 0.675(0.016) & 0.493(0.020) \\
 &  & 3000 & 0.735(0.037) & \textbf{0.788}(0.014) & 0.780(0.014) & 0.680(0.014) & 0.495(0.016) \\
\cmidrule(l){2-8}
 & \multirow[t]{3}{*}{Cosine} & 1000 & 0.323(0.034) & 0.906(0.010) & 0.900(0.010) & 0.899(0.009) & \textbf{0.935}(0.003) \\
 &  & 2000 & 0.350(0.024) & 0.906(0.008) & 0.899(0.008) & 0.898(0.008) & \textbf{0.934}(0.004) \\
 &  & 3000 & 0.369(0.023) & 0.908(0.007) & 0.902(0.007) & 0.900(0.006) & \textbf{0.935}(0.004) \\
\midrule
\multirow[t]{9}{*}{III} & \multirow[t]{3}{*}{GRB} & 1000 & 0.834(0.020) & \textbf{0.965}(0.009) & 0.933(0.022) & 0.834(0.013) & 0.800(0.018) \\
 &  & 2000 & 0.858(0.015) & \textbf{0.979}(0.005) & 0.964(0.011) & 0.834(0.011) & 0.802(0.012) \\
 &  & 3000 & 0.869(0.009) & \textbf{0.982}(0.005) & 0.973(0.009) & 0.832(0.011) & 0.799(0.012) \\
\cmidrule(l){2-8}
 & \multirow[t]{3}{*}{BMC} & 1000 & 0.898(0.030) & \textbf{0.973}(0.004) & 0.935(0.046) & 0.869(0.014) & 0.810(0.045) \\
 &  & 2000 & 0.923(0.007) & \textbf{0.985}(0.002) & 0.975(0.004) & 0.874(0.011) & 0.810(0.051) \\
 &  & 3000 & 0.928(0.006) & \textbf{0.987}(0.001) & 0.982(0.002) & 0.876(0.009) & 0.814(0.045) \\
\cmidrule(l){2-8}
 & \multirow[t]{3}{*}{Cosine} & 1000 & 0.697(0.036) & \textbf{0.922}(0.035) & 0.896(0.063) & 0.754(0.024) & 0.665(0.036) \\
 &  & 2000 & 0.714(0.021) & \textbf{0.951}(0.005) & 0.929(0.040) & 0.749(0.018) & 0.657(0.027) \\
 &  & 3000 & 0.715(0.028) & \textbf{0.957}(0.005) & 0.950(0.005) & 0.751(0.018) & 0.656(0.020) \\
\midrule
\multirow[t]{9}{*}{IV} & \multirow[t]{3}{*}{GRB} & 1000 & 0.736(0.020) & \textbf{0.916}(0.009) & 0.869(0.013) & 0.668(0.049) & 0.272(0.071) \\
 &  & 2000 & 0.763(0.013) & \textbf{0.928}(0.005) & 0.897(0.009) & 0.678(0.029) & 0.309(0.038) \\
 &  & 3000 & 0.774(0.015) & \textbf{0.931}(0.006) & 0.905(0.007) & 0.689(0.027) & 0.300(0.033) \\
\cmidrule(l){2-8}
 & \multirow[t]{3}{*}{BMC} & 1000 & 0.838(0.014) & \textbf{0.936}(0.008) & 0.899(0.013) & 0.818(0.015) & 0.235(0.014) \\
 &  & 2000 & 0.856(0.010) & \textbf{0.931}(0.007) & 0.908(0.009) & 0.819(0.009) & 0.236(0.012) \\
 &  & 3000 & 0.860(0.010) & \textbf{0.933}(0.006) & 0.912(0.008) & 0.819(0.010) & 0.232(0.012) \\
\cmidrule(l){2-8}
 & \multirow[t]{3}{*}{Cosine} & 1000 & 0.348(0.033) & \textbf{0.747}(0.020) & 0.720(0.022) & 0.261(0.052) & 0.141(0.015) \\
 &  & 2000 & 0.378(0.031) & \textbf{0.772}(0.014) & 0.761(0.016) & 0.291(0.042) & 0.144(0.016) \\
 &  & 3000 & 0.391(0.028) & \textbf{0.782}(0.013) & 0.775(0.013) & 0.297(0.029) & 0.141(0.013) \\
\bottomrule
\end{tabular}%
\end{table}

Table~\ref{table1} shows that, except in Model II with the cosine mechanism, GMDD-Net with either kernel embedding or discrete FPCA achieves the highest mean distance correlation. These comparisons indicate that the encoder--network architecture captures more response-relevant information than the competing RKHS methods in the settings considered. The network methods also offer a computational advantage over the competing RKHS methods, and this advantage increases with sample size: at $n=3000$, the average runtimes of f-GSIR and f-GSAVE are approximately 2.48 and 4.31 times that of kernel embedding with GMDD-Net, respectively. We notice that in model I, the mean distance correlations will not improve as $n$ increases. Model I is relatively simple, and the method may already be close to its performance plateau under the current settings with 1000 training samples. Consequently, increasing the sample size may yield only limited improvements. The slight declines observed in this setting may be related to the optimization process and differences between the training objective and the evaluation metric.

Appendix B.2 in the Supplement also reports results measured by the multiple correlation of multivariate rank (MCMR), the criterion used by \cite{LiSong2017}.

\subsection{Metric-space response}
We next retain the functional predictor mechanisms and encoder choices while replacing the Euclidean response by an object in a metric space. We consider the following three models:
\begin{itemize}
\item[V:] $Y$ is a random normal distribution with quantile function 
\begin{equation*}
Q_Y(t)=\mu_Y+\sigma_Y\Phi^{-1}(t),
\end{equation*}
where $\Phi(\cdot)$ is the standard normal cumulative distribution function, $\mu_Y\sim N(0.1\langle X,X\rangle,0.1^2)$, and $\sigma_Y=|\sin(\langle b_1,X\rangle+\langle b_2,X\rangle)|$.
\item[VI:] $Y=\exp(D+0.1E)$ is a $2\times 2$ SPD matrix, where $\exp$ denotes the matrix exponential,
\begin{equation*}
D=\left(
    \begin{array}{cc}
    1 & \quad\rho \\
    \rho & \quad 1
    \end{array}
\right), \quad \rho=\sin\{0.5\pi(\langle X,X\rangle+\langle b_1,X\rangle-1)\},
\end{equation*}
and $E$ is a $2\times2$ symmetric random matrix with independent $N(0,1)$ diagonal entries and $N(0,0.5^2)$ off-diagonal entries.
\item[VII:] $Y=\mathrm{Exp}_pv$ is a random point on the two-dimensional unit sphere. Here $p=(0,0,1)$, $\mathrm{Exp}_p$ is the Riemannian exponential map at $p$, and $v$ is a tangent vector at $p$ given by
\begin{equation*}
v=\Big(e^{0.9\langle b_1,X\rangle}\sin(\langle X,X\rangle)+\epsilon_1,e^{0.9\langle b_2,X\rangle}\cos(\langle X,X\rangle)+\epsilon_2,0\Big)^\t, \quad\epsilon_1,\epsilon_2\sim N(0,0.1^2).
\end{equation*}
\end{itemize}

We combine the three encoders with FCCov-Net and compare them with f-GSIR and f-GSAVE. Since the latter two methods are not tailored for metric-space responses, we transform $Y$ into Euclidean vectors: in model V, form the quantile values into a vector; in model VI, stack the lower-triangle entries into a vector; and in model VII, use the three Cartesian coordinates directly. Such comparison is not entirely fair, so we only present the results of f-GSIR and f-GSAVE as a benchmark. All settings are the same to Section~\ref{simu1} and the results of distance correlation are listed in Table~\ref{table2}.
\begin{table}
\centering
\caption{Mean (standard deviation) of distance correlations over 100 repetitions for three encoders combined with \textbf{FCCov-Net} and for the nested RKHS methods.}
\label{table2}
\begin{tabular}{@{}llcccccc@{}}
\toprule
Model & $X$ generator & $n$ & \shortstack{Fourier\\Expansion} & \shortstack{Kernel\\Embedding} & \shortstack{Discrete\\FPCA} & f-GSIR & f-GSAVE \\
\midrule
\multirow{9}{*}{V} & \multirow{3}{*}{GRB} & 1000 & 0.591\,(0.034) & \textbf{0.786}\,(0.069) & 0.648\,(0.044) & 0.592\,(0.030) & 0.601\,(0.030) \\
 &  & 2000 & 0.595\,(0.034) & \textbf{0.918}\,(0.021) & 0.704\,(0.061) & 0.596\,(0.028) & 0.599\,(0.029) \\
 &  & 3000 & 0.605\,(0.030) & \textbf{0.940}\,(0.008) & 0.826\,(0.086) & 0.602\,(0.027) & 0.599\,(0.024) \\
\cmidrule(lr){2-8}
 & \multirow{3}{*}{BMC} & 1000 & 0.363\,(0.025) & \textbf{0.404}\,(0.025) & 0.358\,(0.029) & 0.380\,(0.023) & 0.368\,(0.021) \\
 &  & 2000 & 0.378\,(0.029) & \textbf{0.413}\,(0.031) & 0.378\,(0.028) & 0.378\,(0.025) & 0.364\,(0.024) \\
 &  & 3000 & 0.392\,(0.023) & \textbf{0.424}\,(0.021) & 0.397\,(0.024) & 0.386\,(0.021) & 0.370\,(0.018) \\
\cmidrule(lr){2-8}
 & \multirow{3}{*}{Cosine} & 1000 & 0.208\,(0.041) & \textbf{0.851}\,(0.032) & 0.786\,(0.041) & 0.615\,(0.036) & 0.574\,(0.034) \\
 &  & 2000 & 0.252\,(0.043) & \textbf{0.873}\,(0.012) & 0.833\,(0.027) & 0.626\,(0.028) & 0.577\,(0.019) \\
 &  & 3000 & 0.286\,(0.041) & \textbf{0.874}\,(0.015) & 0.859\,(0.015) & 0.633\,(0.030) & 0.578\,(0.022) \\
\midrule
\multirow{9}{*}{VI} & \multirow{3}{*}{GRB} & 1000 & 0.519\,(0.027) & \textbf{0.789}\,(0.025) & 0.597\,(0.038) & 0.440\,(0.028) & 0.185\,(0.034) \\
 &  & 2000 & 0.558\,(0.024) & \textbf{0.851}\,(0.015) & 0.721\,(0.026) & 0.445\,(0.026) & 0.196\,(0.027) \\
     &  & 3000 & 0.583\,(0.027) & \textbf{0.876}\,(0.017) & 0.805\,(0.019) & 0.458\,(0.032) & 0.189\,(0.028) \\
\cmidrule(lr){2-8}
     & \multirow{3}{*}{BMC} & 1000 & 0.526\,(0.032) & \textbf{0.872}\,(0.015) & 0.752\,(0.035) & 0.417\,(0.037) & 0.099\,(0.039) \\
     &  & 2000 & 0.611\,(0.024) & \textbf{0.917}\,(0.010) & 0.860\,(0.021) & 0.439\,(0.030) & 0.092\,(0.035) \\
     &  & 3000 & 0.626\,(0.022) & \textbf{0.926}\,(0.009) & 0.906\,(0.012) & 0.439\,(0.023) & 0.087\,(0.024) \\
\cmidrule(lr){2-8}
 & \multirow{3}{*}{Cosine} & 1000 & 0.224\,(0.035) & \textbf{0.797}\,(0.021) & 0.670\,(0.041) & 0.462\,(0.028) & 0.334\,(0.115) \\
 &  & 2000 & 0.256\,(0.031) & \textbf{0.850}\,(0.016) & 0.795\,(0.028) & 0.453\,(0.029) & 0.328\,(0.104) \\
 &  & 3000 & 0.279\,(0.026) & \textbf{0.868}\,(0.013) & 0.840\,(0.020) & 0.455\,(0.028) & 0.328\,(0.110) \\
\midrule
\multirow{9}{*}{VII} & \multirow{3}{*}{GRB} & 1000 & 0.634\,(0.013) & \textbf{0.738}\,(0.016) & 0.702\,(0.015) & 0.725\,(0.013) & 0.658\,(0.032) \\
 &  & 2000 & 0.649\,(0.018) & \textbf{0.745}\,(0.021) & 0.730\,(0.021) & 0.730\,(0.012) & 0.658\,(0.025) \\
 &  & 3000 & 0.654\,(0.013) & \textbf{0.750}\,(0.014) & 0.742\,(0.014) & 0.733\,(0.007) & 0.658\,(0.019) \\
\cmidrule(lr){2-8}
 & \multirow{3}{*}{BMC} & 1000 & 0.720\,(0.019) & \textbf{0.797}\,(0.015) & 0.769\,(0.015) & 0.734\,(0.020) & 0.597\,(0.031) \\
 &  & 2000 & 0.739\,(0.015) & \textbf{0.802}\,(0.013) & 0.793\,(0.012) & 0.739\,(0.016) & 0.594\,(0.028) \\
 &  & 3000 & 0.744\,(0.012) & \textbf{0.800}\,(0.014) & 0.796\,(0.013) & 0.739\,(0.017) & 0.589\,(0.023) \\
\cmidrule(lr){2-8}
 & \multirow{3}{*}{Cosine} & 1000 & 0.376\,(0.023) & \textbf{0.695}\,(0.020) & 0.646\,(0.029) & 0.685\,(0.022) & 0.689\,(0.021) \\
 &  & 2000 & 0.402\,(0.016) & \textbf{0.723}\,(0.015) & 0.690\,(0.021) & 0.688\,(0.012) & 0.689\,(0.011) \\
 &  & 3000 & 0.412\,(0.016) & \textbf{0.736}\,(0.014) & 0.717\,(0.017) & 0.692\,(0.011) & 0.689\,(0.012) \\
\bottomrule
\end{tabular}%
\end{table}

Kernel embedding with FCCov-Net achieves the highest mean distance correlation in all scenarios, while discrete FPCA ranks second in most cases. This highlights again that our methods are of practical utility. Together with the Euclidean-response experiments, these results illustrate how the same functional encoders can support supervised reduction across different response spaces. Results measured by MCMR are also provided in Appendix B.2.

\section{Application}
\begin{figure}
\centering
\includegraphics[width=1\textwidth]{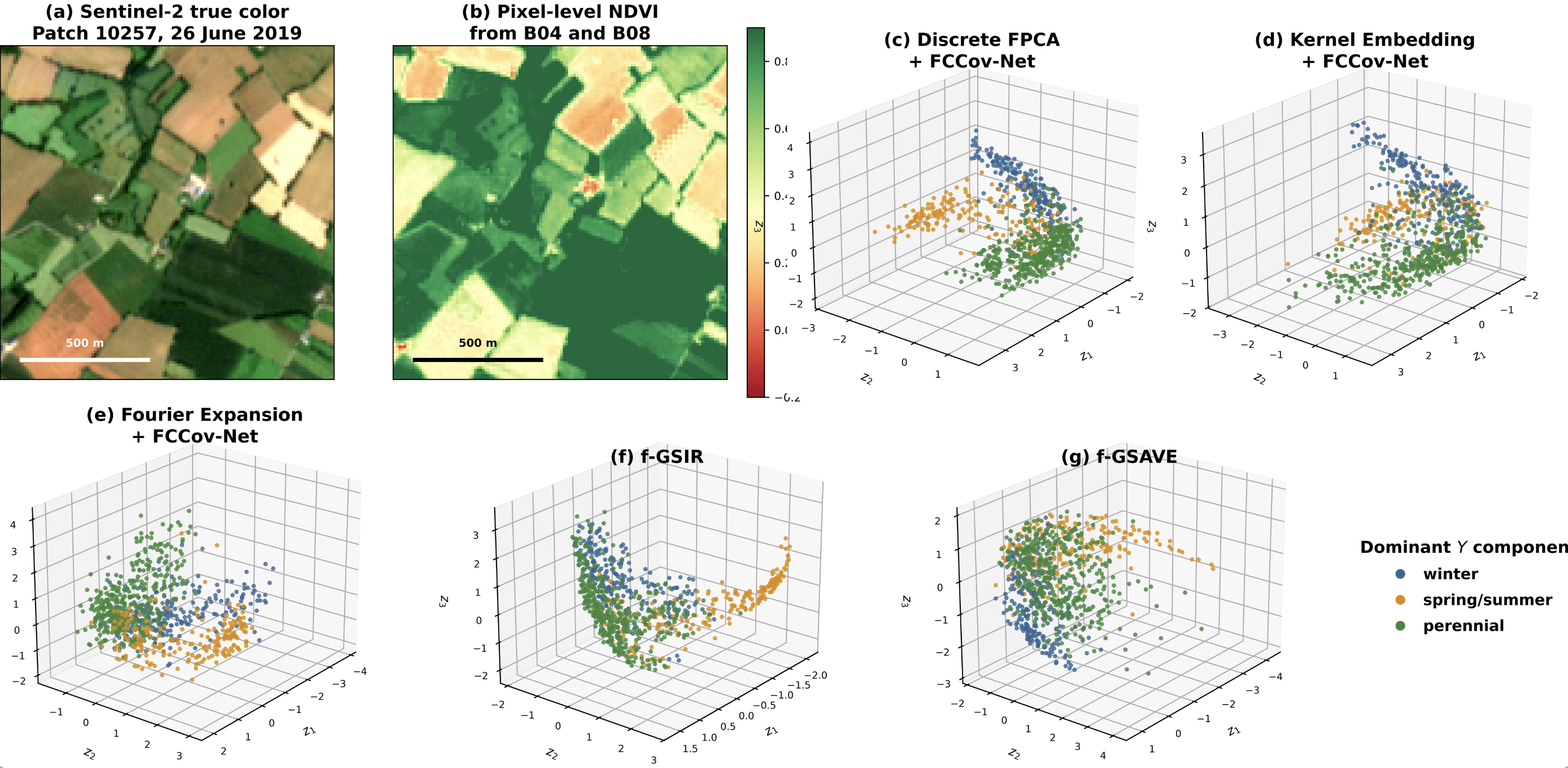}
\caption{(a) Sentinel-2 true-color image of patch 10257 on 26 June 2019, constructed from red (B04), green (B03), and blue (B02) bands; (b) Pixel-level NDVI image of patch 10257 calculated by B04 and B08, from the same date; (c)-(g) 3-dimensional scatter plots of sufficient predictors obtained by three encoders (Discrete FPCA, Kernel Embedding, Fourier Expansion) with FCCov-Net and f-GSIR, f-GSAVE, colored by the largest proportion.}
\label{figure1}
\end{figure}

The PASTIS dataset \cite{pastis} contains 2433 spatial image patches acquired by Sentinel-2, pixel-level agricultural annotations, and five predefined spatial folds. Figure~\ref{figure1}(a) shows a true-color image of one patch. For patch $i$, the satellite data form a four-dimensional array of size $T_i\times 10\times 128\times 128$, where $T_i$ is the number of acquisition dates. The 10 channels correspond to Sentinel-2 bands B02, B03,..., B08, B8A, B11, and B12, and each acquisition is a $128\times 128$ image. Acquisition dates are recorded separately in the metadata; both these dates and $T_i$ may vary across patches. Each patch also has a target array of size $3\times 128\times 128$. We use its first plane, a $128\times128$ semantic crop-label matrix whose integer entries identify the crop or land-cover category of each pixel.

To construct the functional predictor, we use the red band B04 and near-infrared band B08 at each acquisition date. We select these two bands from the 10 available channels because the contrast between red absorption and near-infrared reflection in healthy vegetation makes their normalized difference informative about seasonal crop growth. At each valid pixel $s$, we compute the normalized difference vegetation index (NDVI; \cite{Tucker1979}):
\begin{equation*}
\mathrm{NDVI}_i(t,s)=\frac{\mathrm{B08}_i(t,s)-\mathrm{B04}_i(t,s)}{\mathrm{B08}_i(t,s)+\mathrm{B04}_i(t,s)},
\end{equation*}
and then truncate the result to [-1,1] (see Figure~\ref{figure1}(b) for a pixel-level NDVI image). Then average it over all valid pixels in the $128\times 128$ patch: 
\begin{equation*}
X_i(t)=\frac{1}{\mathcal{V}_i(t)}\sum_{s\in\mathcal{V}_i(t)}\mathrm{NDVI}_i(t,s),
\end{equation*}
where $\mathcal{V}_i(t)$ contains pixels for which both band values are finite and the absolute value of their sum exceeds $10^{-8}$, so that NDVI can be computed reliably. We retain an acquisition only if at least 80\% of the patch pixels and at least 100 pixels in total have valid B04 and B08 values. Observation times are measured in days since September 1, 2018, then mapped to the common calendar domain [0,1]. This places the trajectories on a shared time domain while preserving each curve's original observation locations. The resulting sample contains 856 curves with 38 to 61 observations each.

We construct the response from the $128\times128$ semantic crop-label matrix by consolidating the agricultural labels into three phenological groups: winter annual crops, corresponding to labels $\{2,4,5,10,11\}$; spring/summer annual crops $\{3,6,7,9,13,15,18\}$; and perennial meadow or fodder crops $\{1,8,14,16\}$. Let $C_{ig}$ be the number of pixels in patch $i$ belonging to group $g$. The response is $Y_i=(C_{i1}/\sum_{g=1}^3C_{ig},C_{i2}/\sum_{g=1}^3C_{ig},C_{i3}/\sum_{g=1}^3C_{ig})$.
It is a probability vector on the two-dimensional simplex, providing the distributional response for the functional SDR analysis.

Nonlinear SDR methods are applied to this example when predictor is an irregularly observed function and the response is a probability distribution. Specifically, we want to verify that whether the full NDVI trajectories can be replaced by a low-dimensional vector without losing the information about the compositional response. We first encode each irregular curve into a 10-dimensional vector using three encoders: Fourier basis expansion, kernel embedding and discrete FPCA. Then the encoded vector is passed through an FCCov-Net with architecture 10-64-64-3, producing three-dimensional sufficient predictor $Z_i$. In the FCCov-Net, the Hellinger distance was used to measure dissimilarity on a simplex. $\hat{d}=3$ was chosen from the dimension determination method in Section~\ref{determine}. We present the 3-dimensional scatter plots of $Z_i$ obtained by the three encoders with FCCov-Net and f-GSIR, f-GSAVE in Figure~\ref{figure1}(c)-(g), with points colored according to the category with the largest proportion in $Y_i$. Among the five subfigures, FCCov-Net based on discrete FPCA produces the clearest visual separation among patches with different dominant response categories. This visual pattern is consistent with its superior out-of-fold predictive performance stated below.

Finally, given the estimated coordinates $Z_i$, we predict $Y_i$ using Gaussian-kernel local Fr\'echet regression \cite{PetersonMuller2019}. Prediction is evaluated using the five official PASTIS folds: in each fold, all encoder parameters, FCCov-Net weights, f-GSIR or f-GSAVE, and local-regression tuning parameters are estimated using the other four folds only. The prediction accuracy is reported using out-of-fold Hellinger RMSE and Fr\'echet $R^2$, as shown in Table~\ref{app1}. We also list the dominant-class accuracy of predictions. Formulations of these metrics can be found in Appendix A.2 in the Supplement. We can see that Discrete FPCA with FCCov-Net achieves the lowest Hellinger RMSE and highest Fr\'echet $R^2$ and dominant-class accuracy.

\begin{table}
\centering
\caption{Mean (standard deviation) of prediction metrics across the five PASTIS folds for three encoders combined with \textbf{FCCov-Net} and for the nested RKHS methods.}
\label{app1}
\begin{tabular}{lccc}
\toprule
Method & Hellinger RMSE & Fr\'echet $R^2$ & Dominant-class Accuracy \\
\midrule
Discrete FPCA & 0.164\,(0.014) &0.691\,(0.061) & 0.805\,(0.035) \\
Kernel Embedding & 0.190\,(0.014) & 0.591\,(0.050) & 0.770\,(0.030) \\
Fourier Expansion& 0.198\,(0.012) & 0.550\,(0.064) & 0.740\,(0.019) \\
f-GSIR & 0.193\,(0.013) & 0.575\,(0.046) & 0.760\,(0.013)\\
f-GSAVE & 0.193\,(0.016) & 0.574\,(0.058) & 0.736\,(0.029) \\
\bottomrule
\end{tabular}%
\end{table}

\section{Conclusion}

We have developed a modular framework for functional nonlinear SDR by combining presmoothing encoders with MDD-form networks for Euclidean and metric-space-valued responses. Kernel embedding and discrete FPCA exploit input smoothness through deterministic function-space conditions and covariance spectral decay, respectively. {Our theory quantifies how approximation, statistical estimation, and errors from discrete noisy observations and encoder estimation contribute to the final error rate.} Simulations show gains in accuracy and computational efficiency. Future work includes data-driven selection of smoothing parameters, encoder dimension, and network complexity with guarantees that adapt to unknown input regularity. Another direction is to develop response-informed encoders jointly with the SDR network and analyze their estimation error.

\bibliographystyle{chicago} 
\bibliography{main.bib}   

\begin{thebibliography}{}

\bibitem[\protect\citeauthoryear{Anandkumar, Azizzadenesheli, Bhattacharya, Kovachki, Liu, Liu, and Stuart}{Anandkumar et~al.}{2020}]{Ana2020}
Anandkumar, A., K.~Azizzadenesheli, K.~Bhattacharya, N.~Kovachki, Z.~Liu, B.~Liu, and A.~Stuart (2020).
\newblock Neural operator: Graph kernel network for partial differential equations.
\newblock {\em ICLR 2020 Workshop on Integration of Deep Neural Models and Differential Equations\/}.

\bibitem[\protect\citeauthoryear{Bach}{Bach}{2017}]{Bach2017}
Bach, F. (2017).
\newblock On the equivalence between kernel quadrature rules and random feature expansions.
\newblock {\em Journal of Machine Learning Research\/}~{\em 18}, 714--751.

\bibitem[\protect\citeauthoryear{Chen and Chen}{Chen and Chen}{1995}]{ChenChen1995}
Chen, T. and H.~Chen (1995).
\newblock Universal approximation to nonlinear operators by neural networks with arbitrary activation functions and its application to dynamical systems.
\newblock {\em IEEE Transactions on Neural Networks\/}~{\em 6}, 911--917.

\bibitem[\protect\citeauthoryear{Chen, Jiao, Qiu, and Yu}{Chen et~al.}{2024}]{Chen2024}
Chen, Y., Y.~Jiao, R.~Qiu, and Z.~Yu (2024).
\newblock Deep nonlinear sufficient dimension reduction.
\newblock {\em The Annals of Statistics\/}~{\em 52}, 1201--1226.

\bibitem[\protect\citeauthoryear{Cheng, Fan, Song, and Zhou}{Cheng et~al.}{2026}]{FanZhou2026}
Cheng, K., J.~Fan, L.~Song, and D.-X. Zhou (2026).
\newblock Learning fr\'echet differentiable operators via prespecified neural operators.
\newblock {\em Applied and Computational Harmonic Analysis\/}~{\em 84}, 101878.

\bibitem[\protect\citeauthoryear{Cook and Weisberg}{Cook and Weisberg}{1991}]{CookWeisberg1991}
Cook, R.~D. and S.~Weisberg (1991).
\newblock Discussion of sliced inverse regression for dimension reduction.
\newblock {\em Journal of the American Statistical Association\/}~{\em 86}, 328--332.

\bibitem[\protect\citeauthoryear{Ferr\'e}{Ferr\'e}{1998}]{Ferre1998}
Ferr\'e, L. (1998).
\newblock Determining the dimension in sliced inverse regression and related methods.
\newblock {\em Journal of the American Statistical Association\/}~{\em 93}, 132--140.

\bibitem[\protect\citeauthoryear{Ferr\'e and Yao}{Ferr\'e and Yao}{2003}]{FerreYao2003}
Ferr\'e, L. and A.~F. Yao (2003).
\newblock Functional sliced inverse regression analysis.
\newblock {\em Statistics\/}~{\em 37}, 475--488.

\bibitem[\protect\citeauthoryear{Ferr\'e and Yao}{Ferr\'e and Yao}{2005}]{FerreYao2005}
Ferr\'e, L. and A.-F. Yao (2005).
\newblock Smoothed functional inverse regression.
\newblock {\em Statistica Sinica\/}~{\em 15}, 665--683.

\bibitem[\protect\citeauthoryear{Garnot and Landrieu}{Garnot and Landrieu}{2021}]{pastis}
Garnot, V. S.~F. and L.~Landrieu (2021).
\newblock Panoptic segmentation of satellite image time series with convolutional temporal attention networks.
\newblock {\em IEEE/CVF International Conference on Computer Vision (ICCV)\/}, 4852--4861.

\bibitem[\protect\citeauthoryear{Herrmann, Schwab, and Zech}{Herrmann et~al.}{2024}]{Herrmann2024}
Herrmann, L., C.~Schwab, and J.~Zech (2024).
\newblock Neural and spectral operator surrogates: unified construction and expression rate bounds.
\newblock {\em Advances in Computational Mathematics\/}~{\em 50}.

\bibitem[\protect\citeauthoryear{Hsing and Eubank}{Hsing and Eubank}{2015}]{HsingEubank2015}
Hsing, T. and R.~Eubank (2015).
\newblock {\em Theoretical Foundations of Functional Data Analysis, with an Introduction to Linear Operators}.
\newblock John Wiley \& Sons, Ltd.

\bibitem[\protect\citeauthoryear{Hsing and Ren}{Hsing and Ren}{2009}]{HsingRen2009}
Hsing, T. and H.~Ren (2009).
\newblock An rkhs formulation of the inverse regression dimension reduction problem.
\newblock {\em The Annals of Statistics\/}~{\em 37}, 762--755.

\bibitem[\protect\citeauthoryear{Huang, Jiao, Liao, Liu, and Yu}{Huang et~al.}{2024}]{Huang2024}
Huang, J., Y.~Jiao, X.~Liao, J.~Liu, and Z.~Yu (2024).
\newblock Deep dimension reduction for supervised representation learning.
\newblock {\em IEEE Transactions on Information Theory\/}~{\em 70}, 3583--3598.

\bibitem[\protect\citeauthoryear{Koenderink}{Koenderink}{1984}]{Koen1984}
Koenderink, J.~J. (1984).
\newblock The structure of images.
\newblock {\em Biological Cybernetics\/}~{\em 50}, 363--370.

\bibitem[\protect\citeauthoryear{Kutyniok, Petersen, Raslan, and Schneider}{Kutyniok et~al.}{2022}]{Kutyniok2022}
Kutyniok, G., P.~Petersen, M.~Raslan, and R.~Schneider (2022).
\newblock A theoretical analysis of deep neural networks and parametric pdes.
\newblock {\em Constructive Approximation\/}~{\em 55}, 73--125.

\bibitem[\protect\citeauthoryear{Lanthaler}{Lanthaler}{2023}]{Lanthaler2023}
Lanthaler, S. (2023).
\newblock Operator learning with pca-net: upper and lower complexity bounds.
\newblock {\em Journal of Machine Learning Research\/}~{\em 24}, 1--67.

\bibitem[\protect\citeauthoryear{Lee, Li, and Chiaromonte}{Lee et~al.}{2013}]{LeeLiChia2013}
Lee, K.-Y., B.~Li, and F.~Chiaromonte (2013).
\newblock A general theory for nonlinear sufficient dimension reduction: Formulation and estimation.
\newblock {\em The Annals of Statistics\/}~{\em 41}, 221--249.

\bibitem[\protect\citeauthoryear{Li, Artemiou, and Li}{Li et~al.}{2011}]{LiArteLi2011}
Li, B., A.~Artemiou, and L.~Li (2011).
\newblock Principal support vector machines for linear and nonlinear sufficient dimension reduction.
\newblock {\em The Annals of Statistics\/}~{\em 39}, 3182--3210.

\bibitem[\protect\citeauthoryear{Li and Song}{Li and Song}{2017}]{LiSong2017}
Li, B. and J.~Song (2017).
\newblock Nonlinear sufficent dimension reduction for functional data.
\newblock {\em The Annals of Statistics\/}~{\em 45}, 1059--1095.

\bibitem[\protect\citeauthoryear{Li and Wang}{Li and Wang}{2007}]{LiWang2007}
Li, B. and S.~Wang (2007).
\newblock On directional regression for dimension reduction.
\newblock {\em Journal of the American Statistical Association\/}~{\em 102}, 997--1008.

\bibitem[\protect\citeauthoryear{Li, Zha, and Chiaromonte}{Li et~al.}{2005}]{LiZhaChia2005}
Li, B., H.~Zha, and F.~Chiaromonte (2005).
\newblock Contour regression: a general approach to dimension reduction.
\newblock {\em The Annals of Statistics\/}~{\em 33}, 1580--1616.

\bibitem[\protect\citeauthoryear{Li}{Li}{1991}]{Li1991}
Li, K.-C. (1991).
\newblock Sliced inverse regression for dimension reduction.
\newblock {\em Journal of the American Statistical Association\/}~{\em 86}, 316--327.

\bibitem[\protect\citeauthoryear{Li, Ke, Yin, and Yu}{Li et~al.}{2023}]{Lilu2023}
Li, L., C.~Ke, X.~Yin, and Z.~Yu (2023).
\newblock Generalized martingale difference divergence: detecting conditional mean independence with applications in variable selection.
\newblock {\em Computational Statistics and Data Analysis\/}~{\em 180}.

\bibitem[\protect\citeauthoryear{Lin and Yao}{Lin and Yao}{2019}]{LinYao2019}
Lin, Z. and F.~Yao (2019).
\newblock Intrinsic riemannian functional data analysis.
\newblock {\em The Annals of Statistics\/}~{\em 47}, 3533--3577.

\bibitem[\protect\citeauthoryear{Liu, Yang, Chen, Zhao, and Liao}{Liu et~al.}{2024}]{Liu2023}
Liu, H., H.~Yang, M.~Chen, T.~Zhao, and W.~Liao (2024).
\newblock Deep nonparametric estimation of operators between infinite dimensional spaces.
\newblock {\em Journal of Machine Learning Research\/}~{\em 25}, 1--67.

\bibitem[\protect\citeauthoryear{Lu, Jin, G.~Pang, and Karniadakis}{Lu et~al.}{2021}]{Lu2021}
Lu, L., P.~Jin, Z.~Z. G.~Pang, and G.~E. Karniadakis (2021).
\newblock Learning nonlinear operators via deeponet based on the universal approximation theorem of operators.
\newblock {\em Nature Machine Intelligence\/}~{\em 3}, 218--229.

\bibitem[\protect\citeauthoryear{Ma and Zhu}{Ma and Zhu}{2013}]{MaZhu2013}
Ma, Y. and L.~Zhu (2013).
\newblock Efficient estimation in sufficient dimension reduction.
\newblock {\em The Annals of Statistics\/}~{\em 41}, 250--268.

\bibitem[\protect\citeauthoryear{Ma and Zhu}{Ma and Zhu}{2014}]{MaZhu2014}
Ma, Y. and L.~Zhu (2014).
\newblock On estimation efficiency of the central mean subspace.
\newblock {\em Journal of the Royal Statistical Society: Series B (Statistical Methodology)\/}~{\em 76}, 885--901.

\bibitem[\protect\citeauthoryear{Niu, Cheung, Lin, Dai, Lawrence, and Dunson}{Niu et~al.}{2019}]{Niuetal2019}
Niu, M., P.~Cheung, L.~Lin, Z.~Dai, N.~Lawrence, and D.~Dunson (2019).
\newblock Intrinsic gaussian processes on complex constrained domains.
\newblock {\em Journal of the Royal Statistical Society Series B: Statistical Methodology\/}~{\em 81}, 603--627.

\bibitem[\protect\citeauthoryear{Peterson and M\"uller}{Peterson and M\"uller}{2019}]{PetersonMuller2019}
Peterson, A. and H.-G. M\"uller (2019).
\newblock Fr\'echet regression for random objects with euclidean predictors.
\newblock {\em The Annals of Statistics\/}~{\em 47}, 691--719.

\bibitem[\protect\citeauthoryear{Schott}{Schott}{1994}]{Schott1994}
Schott, J.~R. (1994).
\newblock Determining the dimensionality in sliced inverse regression.
\newblock {\em Journal of the American Statistical Association\/}~{\em 89}, 141--148.

\bibitem[\protect\citeauthoryear{Schwab and Zech}{Schwab and Zech}{2019}]{Schwab2019}
Schwab, C. and J.~Zech (2019).
\newblock Deep learning in high dimension: Neural network expression rates for generalized polynomial chaos expansions in uq.
\newblock {\em Analysis and Applications\/}~{\em 17}, 19--55.

\bibitem[\protect\citeauthoryear{Shao, Lin, and Yao}{Shao et~al.}{2022}]{ShaoLinYao2022}
Shao, L., Z.~Lin, and F.~Yao (2022).
\newblock Intrinsic riemannian functional data analysis for sparse longitudinal observations.
\newblock {\em The Annals of Statistics\/}~{\em 50}, 1696--1721.

\bibitem[\protect\citeauthoryear{Shao and Zhang}{Shao and Zhang}{2014}]{ShaoZhang2014}
Shao, X. and J.~Zhang (2014).
\newblock Martingale difference correlation and its use in high-dimensional variable screening.
\newblock {\em Journal of the American Statistical Association\/}~{\em 109}, 1302--1318.

\bibitem[\protect\citeauthoryear{Shi, Fan, Song, Zhou, and Suykens}{Shi et~al.}{2025}]{Shi2025}
Shi, Z., J.~Fan, L.~Song, D.-X. Zhou, and J.~A. Suykens (2025).
\newblock Nonlinear functional regression by functional deep neural network with kernel embedding.
\newblock {\em Journal of Machine Learning Research\/}~{\em 26}, 1--49.

\bibitem[\protect\citeauthoryear{Smola, Gretton, Song, and Sch\"olkopf}{Smola et~al.}{2007}]{Smola2007}
Smola, A., A.~Gretton, L.~Song, and B.~Sch\"olkopf (2007).
\newblock {\em A Hilbert space embedding for distributions}, pp.\  13--31.
\newblock Springer.

\bibitem[\protect\citeauthoryear{Tang and Li}{Tang and Li}{2026}]{TangLi2026}
Tang, Y. and B.~Li (2026).
\newblock Belted and ensembled neural network for linear and nonlinear sufficient dimension reduction.
\newblock {\em Journal of the American Statistical Association\/}~{\em 121}, 1244--1255.

\bibitem[\protect\citeauthoryear{Tucker}{Tucker}{1979}]{Tucker1979}
Tucker, C.~J. (1979).
\newblock Red and photographic infrared linear combinations for monitoring vegetation.
\newblock {\em Remote Sensing of Environment,\/}~{\em 8}, 127--150.

\bibitem[\protect\citeauthoryear{Wang and Zhong}{Wang and Zhong}{2025}]{JLQi2025}
Wang, J.-L. and Q.~Zhong (2025).
\newblock Joint modeling of longitudinal and survival data.
\newblock {\em Annual Review of Statistics and Its Application\/}~{\em 12}, 449--476.

\bibitem[\protect\citeauthoryear{Xia, Tong, Li, and Zhu}{Xia et~al.}{2002}]{Xia2002}
Xia, Y., H.~Tong, W.~K. Li, and L.-X. Zhu (2002).
\newblock An adaptive estimation of dimension reduction space.
\newblock {\em Journal of the Royal Statistical Society: Series B (Statistical Methodology)\/}~{\em 64}, 363--410.

\bibitem[\protect\citeauthoryear{Yang and Xiang}{Yang and Xiang}{2022}]{YangXiang2022}
Yang, Y. and Y.~Xiang (2022).
\newblock Approximation of functionals by neural network without curse of dimensionality.
\newblock {\em Journal of Machine Learning Research\/}~{\em 1}, 342--372.

\bibitem[\protect\citeauthoryear{Yao and Lei}{Yao and Lei}{2015}]{Yao2015}
Yao, F. and E.~Lei (2015).
\newblock Effective dimension reduction for sparse functional data.
\newblock {\em Biometrika\/}~{\em 102}, 421--437.

\bibitem[\protect\citeauthoryear{Yao, M\"uller, and Wang}{Yao et~al.}{2005}]{Yao2005}
Yao, F., H.-G. M\"uller, and J.-L. Wang (2005).
\newblock Functional data analysis for sparse longitudinal data.
\newblock {\em Journal of the American Statistical Association\/}~{\em 100}, 577--590.

\bibitem[\protect\citeauthoryear{Yarotsky}{Yarotsky}{2018}]{Yaro2018}
Yarotsky, D. (2018).
\newblock Optimal aproximation of continnuous functions by very deep relu networks.
\newblock {\em Conference on Learning Theory\/}, 639--649.

\bibitem[\protect\citeauthoryear{Yuan, Wang, and Yu}{Yuan et~al.}{2026}]{Yuan2026}
Yuan, H., C.~D. Wang, and Z.~Yu (2026).
\newblock Fr\'echet cumulative covariance net for deep nonlinear sufficient dimension reduction with random objects.
\newblock {\em Journal of the American Statistical Association\/}, 1--85.

\bibitem[\protect\citeauthoryear{Zhang and Wang}{Zhang and Wang}{2016}]{ZhangWang2016}
Zhang, X. and J.-L. Wang (2016).
\newblock From sparse to dense functional data and beyond.
\newblock {\em The Annals of Statistics\/}~{\em 44}, 2281--2321.

\bibitem[\protect\citeauthoryear{Zhong, Mueller, and Wang}{Zhong et~al.}{2022}]{QiWang2022}
Zhong, Q., J.~Mueller, and J.-L. Wang (2022).
\newblock Deep learning for the partially linear cox model.
\newblock {\em The Annals of Statistics\/}~{\em 50}, 1348--1375.

\bibitem[\protect\citeauthoryear{Zhong and Wang}{Zhong and Wang}{2024}]{QiWang2024}
Zhong, Q. and J.-L. Wang (2024).
\newblock Neural networks for partially linear quantile regression.
\newblock {\em Journal of Business \& Economic Statistics\/}~{\em 42}, 603--614.

\bibitem[\protect\citeauthoryear{Zhou, Wei, and Yao}{Zhou et~al.}{2025}]{ZhouWeiYao2025}
Zhou, H., D.~Wei, and F.~Yao (2025).
\newblock Theory of functional principal component analysis for discretely observed data.
\newblock {\em The Annals of Statistics\/}~{\em 53}, 2103--2127.

\bibitem[\protect\citeauthoryear{Zhou, Yao, and Zhang}{Zhou et~al.}{2023}]{ZhouYaoZhang2023}
Zhou, H., F.~Yao, and H.~Zhang (2023).
\newblock Functional linear regression for discretely observed data: from ideal to reality.
\newblock {\em Biometrika\/}~{\em 110}, 381--393.

\end{thebibliography}

\end{document}